\documentclass[twocolumn,twocolappendix]{aastex631}
\usepackage{amsmath}
\usepackage{longtable}
\usepackage{hyperref}

\begin{document}

\title{Revisiting the Perseus Cluster III: Role of Aspherical Explosions on its Chemical Composition and Extension to Metal-Poor Stars and Galaxies}

\shortauthors{Leung, Yerdon, Walther, Nomoto and Simionescu}
\shorttitle{Aspherical Explosion for Perseus Cluster, EMPS and Milky Way Stars}

\author[0000-0002-4972-3803]{Shing-Chi Leung}

\affiliation{Department of Physics, SUNY Polytechnic Institute, 100 Seymour Road, Utica, NY 13502, USA}

\author[0009-0000-5656-9659]{Henry Yerdon}

\affiliation{Department of Electrical and Computer Engineering, SUNY Polytechnic Institute, 100 Seymour Road, Utica, NY 13502, USA}
\affiliation{Department of Computer Science, SUNY Polytechnic Institute, 100 Seymour Road, Utica, NY 13502, USA}
\affiliation{Department of Physics, SUNY Polytechnic Institute, 100 Seymour Road, Utica, NY 13502, USA}

\author[0009-0006-8467-2163]{Seth Walther}

\affiliation{Department of Electrical and Computer Engineering, SUNY Polytechnic Institute, 100 Seymour Road, Utica, NY 13502, USA}
\affiliation{Department of Mathematics, SUNY Polytechnic Institute, 100 Seymour Road, Utica, NY 13502, USA}
\affiliation{Department of Physics, SUNY Polytechnic Institute, 100 Seymour Road, Utica, NY 13502, USA}

\author[0000-0001-9553-0685]{Ken'ichi Nomoto}

\affiliation{Kavli Institute for the Physics and 
Mathematics of the Universe (WPI), The University 
of Tokyo Institutes for Advanced Study, The 
University of Tokyo, Kashiwa, Chiba 277-8583, Japan}

\author[0000-0002-9714-3862]{Aurora Simionescu}

\affiliation{SRON Netherlands Institute for Space Research, Niels Bohrweg 4, 2333 CA Leiden, The Netherlands}

\affiliation{Kavli Institute for the Physics and 
Mathematics of the Universe (WPI), The University 
of Tokyo Institutes for Advanced Study, The 
University of Tokyo, Kashiwa, Chiba 277-8583, Japan}

\affiliation{Leiden Observatory, Leiden University, PO Box 9513, 2300 RA Leiden, The Netherlands}

\correspondingauthor{Shing-Chi Leung}
\email{leungs@sunypoly.edu}

\date{\today}

\newcommand{\red}[1]{\textcolor{red}{#1}}

\submitjournal{ApJ}
\received{Nov 10 2025}
\revised{Feb 5 2026}
\accepted{Feb 7 2026}
\published{Apr 7 2026}

\begin{abstract}

The Perseus Cluster has been precisely measured by the legacy Hitomi telescope on the Si-group (Si, S, Ar, Ca) and Fe-group elements (Cr, Mn, Ni). These element abundance ratios provide insight into the typical behaviour of supernovae. In Paper II, we presented new massive star explosion models at various metallicity, assuming spherical explosions. We show that while the fitting is improved, some features (e.g., Ni/Fe) remain to be improved. In this article, we extend our calculation to an aspherical explosion using the jet-induced explosion mechanism. The detailed pre- and post-explosion chemical profiles are calculated with a large post-processing network to capture the production of odd-number elements (V, Mn, Cu) and iron-group elements. We further explore how the jet-driven explosions create the diversity of models which could be compatible with the observed diversity in terms of $^{56}$Ni-mass vs ejecta mass, Ti-V relation, and stellar abundances. Finally, we apply the new collapsar models in the Galactic Chemical Evolution context. We study how the galactic stars, including the Zn-enriched star HE 1327-2326, can put constraints on the relative rates of collapsar and some of its model parameters. We show that collapsar could lead to significant changes in some elements, e.g., Zn. Our study shows that the collapsar is a necessary component to explain multiple elemental trends observed in the Milky Way Galaxy.

\end{abstract}



%\maketitle

%%%%%%%%%%%%%%%%%%%%%%%%%%%%%%%%%%%%%%%%%%%%%%%%%%%%%%%%%%%%%%%%%%%%%%%%%
%%%%%%%%%%%%%%%%%%%%%%%%%%%%%%%%%%%%%%%%%%%%%%%%%%%%%%%%%%%%%%%%%%%%%%%%%

\section{Introduction}

\subsection{Inspiration from Galactic Observations}

Supernovae synthesize metal and distribute them in the universe, enriching the metal content in later generations of stars. Galaxies and Galactic Clusters are ideal test cases to examine the general behaviours of the supernova models and their populations. 

The intrinsic scatters of chemical elements (e.g., [$\alpha$/Fe]) in the galactic stars may reveal their local supernova history \citep{Vicenzo2021GalaxyStars}. These ratios are similarly observed in galactic groups, which leads to constrains on the relative fraction of Type Ia supernovae \citep{Sarkar2022SuzakuGalaxyGroups}.

Some specific elements, such as [Ni/Fe], can provide strong indication on the supernova demography such as the fraction of SNe Ia evolved from sub-Chandrasekhar mass WDs in some dwarf spheroidal galaxies and the delay time \citep{Kirby2019DSphNi}. These measurements demonstrated the possibilities to use these gigantic objects to develop generic constraints on the supernovae explosion models.

\subsection{Aspherical Explosion Models}

Typical massive star explosion models in the literature assume a spherical explosion. It assumes that when the iron core in those stars reaches the Chandrasekhar mass, gravitational instability triggers its spontaneous collapse, leading to the formation of compact objects and the subsequent explosion \citep[see reviews, e.g.,][]{Janak2007CCSNReview, Mueller2020CCSNReview}. The explosion is frequently represented by a mass cut of the Fe-core and a deposition of thermal or kinetic energy in the innermost region. The spherical models are extensively developed using different numerical algorithms and nuclear reaction networks to compute their explosive nucleosynthesis \citep[see e.g.,][]{Heger2002ZeroMet,Chieffi2004CCSN,Tominaga2007,Nomoto2013ARAA}.  

The discovery of the long Gamma-Ray Burst (GRB) GRB 980425 \citep{Soffitta1998GRB980425} and its associated optical afterglow at 0.9 days after GRB in the optical and X-ray band as the SN 1998bw \citep{Galama1998SN1998bw} has led to strong evidence that some GRB has a supernova origin \citep{Galama1999SN1998bw}. Its very high luminosity and broad spectral lines in C+O suggested a very energetic explosion in the order of $10^{52}$ erg \citep{Iwamoto1999SN1998bw, Woosley1999SN1998bw}. The derived $^{56}$Ni $\sim 0.4-0.7~M_{\odot}$ from the light curve is also $\sim 10$ times higher than typical core-collapse SNe, e.g., SN 1987A. The new explosion model, noted as ``hypernova'', is introduced to describe phenomenologically the rare and energetic explosion of this kind \citep{Nomoto2001SN1998bw}. 

The spherical models have difficulties explaining all observational features. They require an energy budget $\sim 10^{52}$ erg, which is theoretically difficult to realize, and the model cannot directly match the asymmetric nature of the GRB \citep{Hoeflich1999SN1998bw}. A mediating model is the collapse of a rapidly rotating star at low metallicity, known as the collapsar \citep{Woosley1999Colapsar}. Collapsar focuses on low-metallicity stars, where mass loss during the main-sequence phase is suppressed. This preserves a massive hydrogen core and keeps the angular momentum of the star. During collapse, the conservation of angular momentum supports the formation of a rapidly rotating compact object, where the accretion flow around this object could trigger an outburst by magneto-rotational instability \citep[MRI, ][]{Balbus1991MRI, Balbus2003MRI}. The production of a LeBlanc-Wilson MHD jet \citep{LeBlamc1970Jet}, typically along the rotation axis, is further accelerated by the density gradient \citep{Cen1998GRB, Wheeler2000Jets}. It results in a highly asymmetric explosion \cite[see early realization in, e.g., ][]{Khokhlov1999JetSN}. The breakout of the jet at the stellar envelope could lead to a cone-shaped outburst, which aligns with the directional dependence of the GRB \citep{MacFadyen1999Collapsar, Zhang2003Jet}.

The collapsar jet sets the framework for the later asymmetric explosion. Early asymmetric models focused on Type II supernovae for the application of aspherical SN remnant \citep[e.g., Cassiopeia A, see ][]{Nagataki1997AsymTypeII}, but were later applied to H-poor stars for connecting with GRB \citep{Zhang2006GRB}. The aspherical explosion provides a natural mechanism for the mixing of matter \citep{Nagataki1998JetMixing}. The cone-shaped structure allows energy to be focused in a small open angle, which favours the shock to propagate \citep{Zhang2004JetProp}. Note that in 1D spherical simulations, the mixing-and-fallback model is commonly used to mimic the mixing effects \citep{Umeda2002PopIIIStar, Ishigaki2021MWStars, Grimmett2021Jet}. Its $^{56}$Ni production is sensitive to the jet energetics \citep{Nagataki2003JetNuc, Nagataki2006JetHyp}. The jet also creates a much higher entropy zone compared to a spherical explosion for the same explosion energy. This favours the production of heavy Fe-group elements such as scandium, titanium, cobalt, and zinc\citep{Tominaga2009Jet}. Such enhanced productions align with some Zn-enriched metal-poor stars \citep{Cayrel2004FirstStar, Ezzeddine2019ZnEMP}. 

\subsection{Motivation}

The Perseus Cluster measured by the legacy Hitomi telescope has inspired this series of articles. The high-resolution spectrum reported in \cite{Simionescu2019Perseus} showed that most massive star yields in the literature cannot simultaneously explain the measured ratios of Si, S, Ar, and Ca with respect to Fe. Systematic overproduction of Si and S, and underproduction of Ar and Ca, are observed. 

In Paper I \citep{Leung2025Perseus1}, we examined the role of convective mixing parameters (including the mixing length parameter $\alpha$ and the semi-convection parameter $\alpha_{\rm SC}$), and we found the set of parameters that could support higher production of Ar and Ca, without simultaneously creating more Si and S. In that work, we computed the massive star models of $15 - 40 M_{\odot}$ in solar metallicity. 

In Paper II (Leung et al. submitted to ApJ), we further expanded our models to 60 $M_{\odot}$, with a metallicity from 0 to $Z_{\odot}$. We apply the massive star catalogue in the Galactic Chemical Evolution (GCE) code, where we study how the new massive star models change the trend of the chemical elements (Si, S, Ar, Ca, Mn, and Ni) at the galactic level. We also compare the metal production with other massive stars and Type Ia supernova models. While we obtain a good fit for most elements, some systematic mismatch is observed, including the underproduction of S and Mn, and the overproduction of Ni. 

Given that the aspherical explosion is one common representation for massive star explosions, as indicated by various aspherical supernova remnants, our previous results raise the question of whether the jet-triggered explosion can consistently address the remaining mismatch. We will examine the nucleosynthetic yield of the new massive star models assuming aspherical explosions. We include the explosive nucleosynthesis yield in our GCE simulations and ask if the existing mismatch can be eased, where we can derive empirical constraints on the rate of collapsar, and hence its role in the supernova population.

In this article, we will describe the results in two parts: the collapsar simulations and the GCE. In Section \ref{sec:methods}, we briefly describe the numerical methods for solving the aspherical explosion of massive stars. We introduce the models studied, and then we present the explosion dynamics and the associated nucleosynthesis. We also compare the new collapsar models with the characteristic model using other massive star progenitors. We also compare the new model with other progenitor pre-explosion models from the literature. In Section \ref{sec:GCE} we describe the modification of the GCE code to include the chemical yield of the new collapsar models as a new supernova channel. We show the best-fit model, which could best reproduce the chemical trends from stars in the Milky Way, where we discuss in detail the actual evolution of those elements (Si, S, Ar, Ca, Mn, and Ni). We also consider other elements that could constrain the collapsar's rate. In Section \ref{sec:discussion}, we compare the derived best-fit collapsar rates and metallicity with other stellar population models in the literature. Finally, we give our conclusion. 

\section{Collapsar Explosion}
\label{sec:collapsar}

\subsection{Methods}
\label{sec:methods}

We use the two-dimensional special relativistic hydrodynamics code to simulate the very energetic jet outburst of the massive star. The code is built on the non-relativistic Euler equations \citep{Leung2015a}, and extended to model the high velocity jet \citep{Leung2023Jet1, Leung2024Jet2}. The code 
has been extensively used for other types of supernovae, including Type Ia supernovae \citep{Leung2015b, Leung2018Chand, Leung2018SubChand}, electron capture supernovae \citep{Zha2019ECSN, Leung2018ECSN} and accretion induced collapse \citep{Leung2019AIC, Zha2019AIC}.

The code solves the Euler equations in spherical coordinates $(r,\theta)$, namely
\begin{equation}
\frac{\partial {\bf D}}{\partial t} + \nabla \cdot {\bf F} = {\bf S}, 
\end{equation}
where 
\begin{eqnarray}
    {\bf D} = (\rho \Gamma, \rho \Gamma^2 h {\bf v}, \rho \Gamma^2 h - p - \rho \Gamma)^T \\ \nonumber 
    {\bf F} = (\rho \Gamma {\bf v}, \rho \Gamma^2 h {\bf v} {\bf v}, \rho \Gamma^2 h {\bf v} - D {\bf v})^T \\ \nonumber 
    {\bf S} = (0, \rho \nabla \Phi, \rho {\bf v} \cdot \nabla \Phi)^T
\end{eqnarray}
are the conservative mass-energy, momentum and energy flux with $\rho, p, \epsilon, {\bf v}$ being the standard hydrodynamics variables (the density, pressure,
specific internal energy and the velocity of the fluid). $h = 1 + p / \rho + \epsilon$ is the specific enthalpy
of the fluid. $\Phi$ is the gravitational potential which satisfies the Poisson equation $\nabla^2 \Phi = 4 \pi G \rho$. $\Gamma = 1 / \sqrt{1 - {\bf v}^2}$is the Lorentz contraction factor. 

This hydrodynamics component is solved by the 
fifth-order weighted essentially non-oscillatory scheme for spatial discretization \citep{Shu1999} and the five-step third-order non-strong stability preserving Runge-Kutta scheme for the time-discretization \citep{Wang2007}. The simulation grid is fixed at (245,60) but with exponentially increasing size in the radial direction from the mass cut to $\sim 4 \times 10^5$ km. The angular direction has a uniform grid size. 

The microphysics is completed by the \texttt{Helmholtz} equation of state \citep{Timmes1999Helm, Timmes2000Helm} which contains the electron gas at arbitrary relativistic and degenerate levels, nucleons as a classical ideal gas, photon gas, and corrections from electron-positron pairs. To describe the chemical composition, we use the 7-isotope network \citep{Timmes20007iso} which can represent the typical nuclear luminosity of a much larger nuclear network with high accuracy. We also include the electron fraction as an independent scalar field to track the effect of electron capture, which is important for degenerate matter $\rho \sim 10^{8-9}$ gcm$^{-3}$ at high temperature. 

To compute the chemical composition with a large network, we use the tracer particle method \citep{Travaglio2011Tracer}, which records the thermodynamic trajectory of the fluid parcels in the exploding star. We pass the density and temperature as a function of time in the large 495-isotope network (containing isotopes from $^{1}$H to $^{91}$Tc) \texttt{Torch} \citep{Timmes1999Torch} to compute the exact nucleosynthetic yield. 

We refer interested readers to the relevant documentation papers \citep{Leung2015a, Leung2023Jet1} for the detailed numerical tests and characteristic models. 

\subsection{Models}

In this section, we present the collapsar hydrodynamics simulation models computed in this work. 

In Table \ref{tab:progenitor} we list the pre-explosion properties of the massive stars in terms of the He-, C+O-, Si-, and Fe-core mass. In general, the He-core is less than half of the ZAMS mass, where the C+O core basically reaches the surface for star mass $\geq 30~M_{\odot}$. 

In Table \ref{tab:models} we list the collapsar models computed in this work. To facilitate the discussion, the model name MXX-YYY-ZZZ stands for the progenitor of XX $M_{\odot}$ ZAMS stars, with YYY (in percent) of the default jet energy deposition rate (in units of $1.2 \times 10^{53}$ erg s$^{-1}$), and ZZZ (in percent) of the default deposition time (in units of 0.0125 s). The units are chosen to conform with previous works \citep{Tominaga2009Jet, Leung2023Jet1}, and the default model deposits a total energy of $1.5 \times 10^{52}$ erg. In this work, we fix the jet open angle to be 15$^{\circ}$. 

\begin{table}[]
    \centering
    \caption{The pre-supernova models used for the explosion simulations. $M_{\rm ZAMS}$, $M_{\rm He}$, $M_{\rm C}$, $M_{\rm Si}$ and $M_{\rm Fe}$ are the progenitor zero-age main-sequence mass, the He-, C-, Si- and Fe-core masses at the onset of collapse, in units of $M_{\odot}$. }
    \begin{tabular}{c c c c c c}
         \hline
         Progenitor & $M_{\rm ZAMS}$ &  $M_{\rm He}$ & $M_{\rm C}$ & $M_{\rm Si}$ & $M_{\rm Fe}$ \\ \hline
         M20 & 20 & 6.00 & 3.71 & 1.82 & 1.38 \\
         M30 & 30 & 10.88 & 10.17 & 1.97 & 1.51 \\
         M40 & 40 & 16.25 & 15.89 & 2.92 & 1.72 \\ \hline
    \end{tabular}
    
    \label{tab:progenitor}
\end{table}

\begin{table*}[]
    \centering

    \caption{The explosion simulations performed in this work. Progenitor stands for the models in Table \ref{tab:progenitor}. $\dot{E}_{\rm dep}$ is the energy deposition rate by the jet in units of $1.2 \times 10^{53}$ erg s$^{-1}$. $T_{\rm dep}$ is the energy deposition duration in units of 0.125 s. $E_{\rm dep}$ is the total deposited energy in units of $10^{52}$ erg s$^{-1}$. $\theta_{\rm jet}$ is the opening angle of the jet in unit of degrees. $R_{\rm cut}$ is the mass-cut radius in km. $M({\rm ^{56}Ni})$ and $M_{\rm ej}$ are the mass of synthesized $^{56}$Ni by the end of simulations and the ejected mass, in units of $M_{\odot}$. [X/Fe] is the scaled mass fraction of the element X (X = Si, S, Ar, Ca) after all short-term radioactive isotopes have decayed.}
    \begin{tabular}{c c c c c c c c c c c c c}
        \hline
         Model & Progenitor & $\dot{E}_{\rm dep}$ & $T_{\rm dep}$ & $\theta_{\rm jet}$ & $R_{\rm cut}$ & $E_{\rm dep}$ & $M_{\rm ej}$ & $M({\rm ^{56}Ni})$ & [Si/Fe] & [S/Fe] & [Ar/Fe] & [Ca/Fe]\\ \hline
         M20-025-050 & M20 & 0.25 & 0.50 & 15 & 1500 & 0.188 & 3.5916 & 0.0407 & 0.4967 & 0.1638 & 0.0197 & 0.0702 \\
         M20-025-100 & M20 & 0.25 & 1.00 & 15 & 1500 & 0.375 & 3.8802 & 0.0459 & 0.4900 & 0.1640 & 0.0145 & 0.06677 \\
         M20-050-025 & M20 & 0.50 & 0.25 & 15 & 1500 & 0.188 & 4.0362 & 0.0426 & 0.4919 & 0.1108 & -0.0458 & 0.0520 \\
         M20-050-050 & M20 & 0.50 & 0.50 & 15 & 1500 & 0.375 & 4.9254 & 0.0490 & 0.5763 & 0.2626 & 0.1192 & 0.1989 \\
         M20-050-100 & M20 & 0.50 & 1.00 & 15 & 1500 & 0.750 & 5.1606 & 0.0531 & 0.5476 & 0.2190 & 0.0784 & 0.1640 \\
         M20-050-200 & M20 & 0.50 & 2.00 & 15 & 1500 & 1.500 & 5.1930 & 0.0439 & 0.5847 & 0.1482 & -0.0516 & 0.0282 \\
         M20-100-025 & M20 & 1.00 & 0.25 & 15 & 1500 & 0.375 & 5.3742 & 0.0555 & 0.5490 & 0.1929 & 0.0339 & 0.1360 \\
         M20-100-050 & M20 & 1.00 & 0.50 & 15 & 1500 & 0.750 & 5.2692 & 0.0533 & 0.5731 & 0.2352 & 0.0723 & 0.1534 \\
         M20-100-100 & M20 & 1.00 & 1.00 & 15 & 1500 & 1.500 & 5.4276 & 0.0593 & 0.5952 & 0.2859 & 0.1209 & 0.1868 \\
         M20-100-200 & M20 & 1.00 & 2.00 & 15 & 1500 & 3.000 & 5.2182 & 0.0097 & 1.1761 & 0.7112 & 0.4690 & 0.5201 \\
         M20-200-050 & M20 & 2.00 & 0.50 & 15 & 1500 & 1.500 & 5.4654 & 0.0632 & 0.5855 & 0.3073 & 0.1605 & 0.2215 \\
         M20-200-100 & M20 & 2.00 & 1.00 & 15 & 1500 & 3.000 & 5.4996 & 0.0453 & 0.7393 & 0.4324 & 0.2649 & 0.2838 \\ \hline

         M30-025-050 & M30 & 0.25 & 0.50 & 15 & 1500 & 0.188 & 0.3852 & 0.0127 & -0.2129 & -0.3666 & -0.4697 & -0.2826 \\
         M30-025-100 & M30 & 0.25 & 1.00 & 15 & 1500 & 0.375 & 1.2664 & 0.0362 & -0.1286 & -0.2137 & -0.3084 & -0.1989 \\
         M30-050-025 & M30 & 0.50 & 0.25 & 15 & 1500 & 0.188 & 1.2904 & 0.0008 & 0.4979 & 0.1775 & 0.1455 & 0.3569 \\
         M30-050-050 & M30 & 0.50 & 0.50 & 15 & 1500 & 0.375 & 1.7158 & 0.0104 & -0.2615 & -0.5501 & -0.6540 & -0.3507 \\
         M30-050-100 & M30 & 0.50 & 1.00 & 15 & 1500 & 0.750 & 5.1277 & 0.0805 & 0.1413 & 0.0625 & -0.0368 & 0.0668 \\
         M30-050-200 & M30 & 0.50 & 2.00 & 15 & 1500 & 1.500 & 4.8557 & 0.0853 & -0.0041 & -0.0898 & -0.1761 & -0.0409 \\
         M30-100-025 & M30 & 1.00 & 0.25 & 15 & 1500 & 0.375 & 2.8473 & 0.0269 & -0.0655 & -0.3423 & -0.5431 & -0.4636 \\
         M30-100-050 & M30 & 1.00 & 0.50 & 15 & 1500 & 0.750 & 7.4604 & 0.0885 & 0.2498 & 0.1461 & 0.0396 & 0.1182 \\
         M30-100-100 & M30 & 1.00 & 1.00 & 15 & 1500 & 1.500 & 7.3233 & 0.1102 & 0.1489 & 0.0469 & -0.0599 & 0.0347 \\
         M30-100-200 & M30 & 1.00 & 2.00 & 15 & 1500 & 3.000 & 7.5105 & 0.1106 & 0.1701 & 0.0641 & -0.0454 & 0.0489 \\
         M30-200-050 & M30 & 2.00 & 0.50 & 15 & 1500 & 1.500 & 8.9640 & 0.1096 & 0.2898 & 0.1780 & 0.0558 & 0.1250 \\
         M30-200-100 & M30 & 2.00 & 1.00 & 15 & 1500 & 3.000 & 9.0467 & 0.1147 & 0.2733 & 0.1582 & 0.0256 & 0.1006 \\ \hline

         M40-025-050 & M40 & 0.25 & 0.50 & 15 & 1500 & 0.188 & 0.9084 & $<$0.0001 & 1.9862 & 1.5995 & 1.5760 & 1.4027 \\
         M40-025-100 & M40 & 0.25 & 1.00 & 15 & 1500 & 0.375 & 0.6029 & 0.0134  & -0.1045 & -0.1767 & -0.2771 & -0.1399 \\
         M40-050-025 & M40 & 0.50 & 0.25 & 15 & 1500 & 0.188 & 0.2779 & $<$0.0001 & 1.2690 & -0.9623 & -2.2208 & 0.0534 \\
         M40-050-050 & M40 & 0.50 & 0.50 & 15 & 1500 & 0.375 & 0.2356 & 0.0029 & 0.2026 & 0.2650 & 0.2423 & 0.2782 \\
         M40-050-100 & M40 & 0.50 & 1.00 & 15 & 1500 & 0.750 & 1.9760 & 0.0206 & -0.3827 & -0.6261 & -0.6503 & -0.4025 \\
         M40-050-200 & M40 & 0.50 & 2.00 & 15 & 1500 & 1.500 & 1.8249 & 0.0228 & -0.6979 & -0.8181 & -0.8723 & -0.7115 \\
         M40-100-025 & M40 & 1.00 & 0.25 & 15 & 1500 & 0.375 & 0.7296 & 0.0008 & -0.2866 & -2.5238 & -2.8043 & -1.4486 \\
         M40-100-050 & M40 & 1.00 & 0.50 & 15 & 1500 & 0.750 & 2.1710 & 0.0099 & -0.0597 & -0.2095 & -0.2874 & -0.1614 \\
         M40-100-100 & M40 & 1.00 & 1.00 & 15 & 1500 & 1.500 & 6.3684 & 0.0521 & 0.0661 & -0.0175 & -0.0646 & 0.0689 \\
         M40-100-200 & M40 & 1.00 & 2.00 & 15 & 1500 & 3.000 & 7.2914 & 0.0596 & 0.0565 & -0.0626 & -0.1382 & -0.0236 \\
         M40-200-050 & M40 & 2.00 & 0.50 & 15 & 1500 & 1.500 & 10.2879 & 0.0705 & 0.2128 & 0.0960 & 0.0199 & 0.1130 \\
         M40-200-100 & M40 & 2.00 & 1.00 & 15 & 1500 & 3.000 & 11.6870 & 0.0845 & 0.2567 & 0.1544 & 0.0570 & 0.1210 \\ \hline
    \end{tabular}
    
    \label{tab:models}
\end{table*}

For 20 $M_{\odot}$, the explosion is less sensitive to the jet parameter because the binding is small compared to the deposited energy, thus the jet can efficiently disrupt the entire star. However, due to the small Si-core, the generated $^{56}$Ni remains around $0.05~ M_{\odot}$. On the other hand, the jet-dependence of the M30 and M40 series is stronger. The ejecta mass ranged from $<1 M_{\odot}$ in the weaker side, and all the way to $\sim 10~M_{\odot}$ on the other side of the spectrum. The $^{56}$Ni has a spread from $0.01 - 0.1~M_{\odot}$, similar to typical core-collapse supernovae. 

\subsection{Explosion Dynamics}

\begin{figure}
    \centering
    \includegraphics[width=0.5\textwidth]{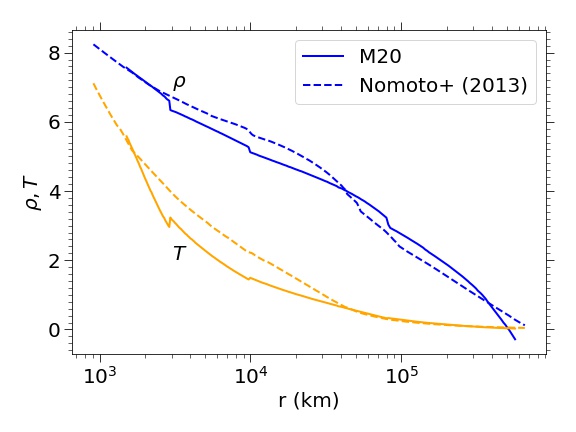}
    \includegraphics[width=0.5\textwidth]{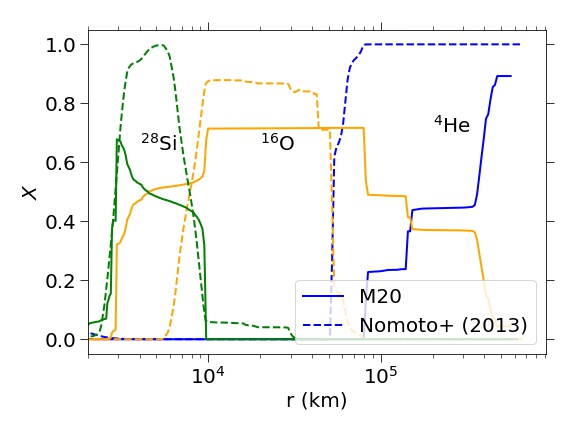}
    \caption{(top panel) The pre-explosion density (blue lines) and temperature (orange lines) profiles for the M40-series. The solid lines correspond to the initial profile used in this work, and the dashed lines correspond to that from \cite{Nomoto2013ARAA} used in \cite{Leung2023Jet1}.
    (bottom panel) The pre-explosion chemical composition profile used for the M40 series (solid lines) and that from \cite{Nomoto2013ARAA}. The $^{28}$Si (green lines), $^{16}$ (orange lines) and $^{4}$He (blue lines) profiles are presented. }
    \label{fig:init_model}
\end{figure}

We plot in \ref{fig:init_model} the initial hydrodynamics (top panel) and chemical composition profiles (bottom panel) of the characteristic model M40-100-100. We also include the previous characteristic model reported in \citep{Leung2023Jet1} for comparison. The two progenitor models are evolved from a 40 $M_{\odot}$ ZAMS star with different stellar evolution codes, but are exploded with the same jet parameters for comparison.

The density and temperature of the two models are approximately the same, except that the upper Si-core and lower O-core in our new model
have a lower temperature and density. The layered structure in the new model shows a clearer step-like transition, which, from the chemical composition file they corresponds to the transition from one shell to another. In terms of the chemical composition, the new model has a more extensive mixing zone, as reflected by a wider layer with multiple elements. For example, some $^{16}$O is mixed in the bottom of $^{4}$He. As discussed in \cite{Leung2025Perseus1}, such mixing near the Si- and O-shells is important to boost the production of Si-group isotopes.  

\begin{figure}
    \centering
    \includegraphics[width=0.5\textwidth]{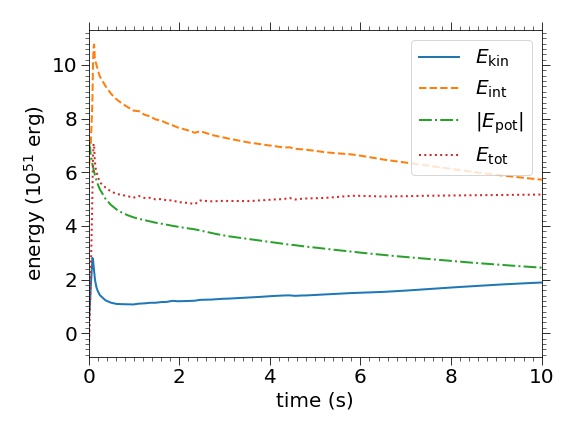}
    \includegraphics[width=0.5\textwidth]{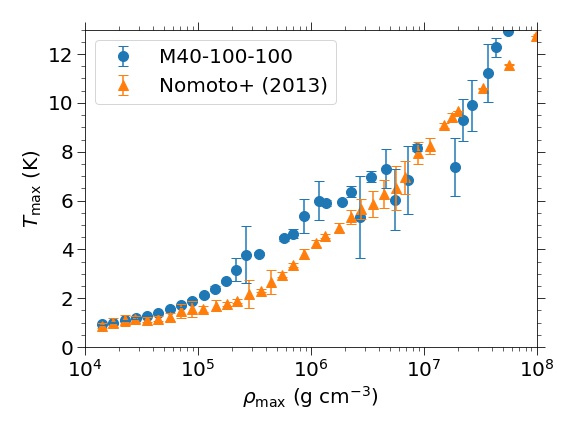}
    \caption{(top panel) The energy evolution of the characteristic model M40-100-100 for the kinetic energy (blue solid line), internal energy (orange dashed line), absolute value of the potential energy (green dot-dashed line) and the total energy (red dotted line). 
    (bottom panel) The statistics of the thermodynamics history of the tracers, showing the tracer maximum temperature and their corresponding density. Each point corresponds to the average $T_{\rm max}$ for each density bin and the error bar represents the standard deviation. The M40-100-100 (blue circles) and N40-1000-1000 \citep[orange triangles, ][]{Leung2023Jet1}. 
    }
    \label{fig:exp_dyn}
\end{figure}

In Figure \ref{fig:exp_dyn} we plot in the upper panel the energy evolution of the characteristic models. The initial jet deposition pumps in a substantial amount of internal energy, which creates a cone-shaped shock front. Such a shock front then later expands during its propagation along the axis. However, the efficiency of the jet deposition is not high. Much of the shocked material falls back at the bottom of the shock, which is reflected by the loss of total energy. The energy fluctuation stabilized at $\sim 0.3$s after the initial deposition. Following the expansion of the shock, the thermal pressure work done is converted into kinetic energy for the outer stellar envelope to expand and escape from the star. 

In the bottom panel, we show the thermodynamic history of the tracers. To construct the plot, we first collect the density of each tracer when they experience the maximum temperature. This determines the levels of nuclear reactions that determine the final chemical composition. Then we bin the quantities by their density, where we compute their average and standard deviation, to characterize the distribution of these tracers. The two models have a qualitatively similar trend, but the new model has generally higher $T_{\rm max}$ between $10^5-5\times10^6$ g cm$^{-3}$. This regime is responsible for incomplete burning, which yields Si-group elements. Between $5 \times 10^6 - 5 \times 10^7$ g cm$^{-3}$, the tracers have a large spread because this corresponds to the mixture which experiences direct energy deposition (within the cone) and experiences only shock-triggered expansion (outside the cone). Then, beyond $\rho_{\rm max} > 5 \times 10^7$ g cm$^{-3}$, tracers in the new models have a higher $T_{\rm peak}$, although tracers with such a high density occupy a very small fraction of the population. From these trends, we expect that the new models generate more Si-group and high entropy isotopes.  

\subsection{Nucleosynthesis}

\begin{figure*}
    \centering
    \includegraphics[width=0.99\linewidth]{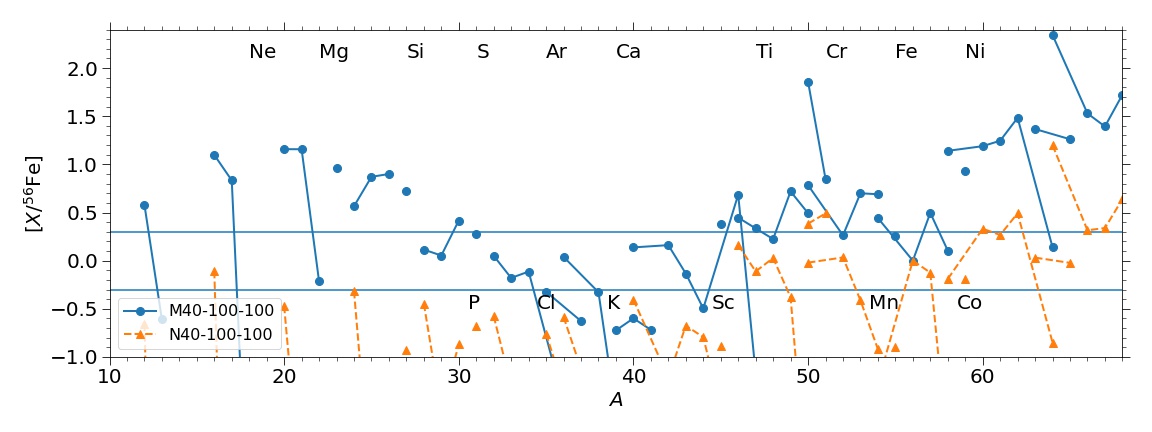}
    \caption{The chemical composition [X$_i$/$^{56}$Fe] for M40-100-100 after all short-lived radioactive isotopes have decayed. The two horizontal lines correspond to two times (upper line) and half (lower line) of the solar value.}
    \label{fig:M40_xiso}
\end{figure*}

\begin{figure}
    \centering
    \includegraphics[width=0.5\textwidth]{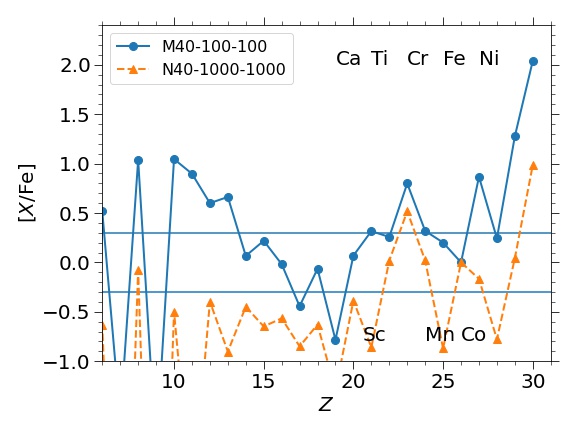}
    \caption{The elemental mass fraction [X/Fe] for M40-100-100.}
    \label{fig:M40_ele}
\end{figure}

In Figure \ref{fig:M40_xiso} we plot the chemical abundance profile [X/$^{56}$Fe]\footnote{[X/$^{56}$Fe] is defined as $\log_{10}$ ((X/$^{56}$Fe)/(X/$^{56}$Fe)$_{\odot}$). Thus, the 0 value means the abundance ratio is the same as the solar composition.} of the characteristic models assuming all short-lived radioactive isotopes have decayed. The two models show very distinctive composition profiles. The new model (M40-100-100) has a very pronounced low-mass elements from O to Si, and near-solar values from Si to Fe, including the odd-number elements, except the super-solar value for V. On the contrary, the old model shows sub-solar values for lower mass elements up to Sc, and near-solar values for Fe-group elements. Both models show super-solar values for Zn. 

In Figure \ref{fig:M40_ele}, we plot the elemental abundance and the comparison of the characteristic models. The super-solar production is more clearly illustrated from Ne to Si and from Fe onwards in the new model. On the other hand, the old model overproduces V and Zn, Fe-group elements (Ti, Cr, Fe, Ni, and Co), agreeing with solar values, and other elements are underproduced. Both models fail to produce enough Cl and K to be compatible with the solar abundance. The overproduction of Zn and other elements in the new characteristic models, as we discuss in a later section, is an important feature that we can adapt to constrain their relative rates. 

\section{Dependence on Explosion Models}

In this section, we examine the mass dependence of the collapsar explosion models. 

\begin{figure}
    \centering
    \includegraphics[width=0.48\textwidth]{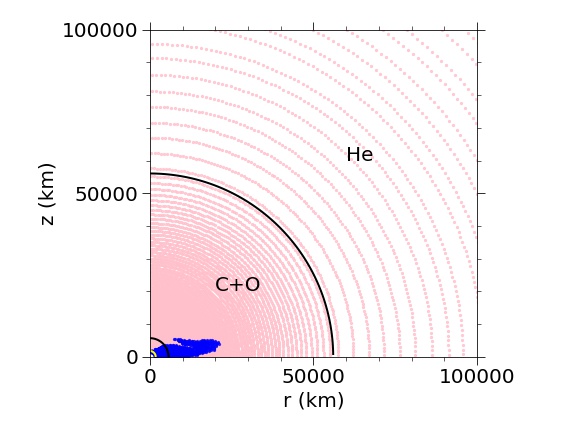}
    \includegraphics[width=0.48\textwidth]{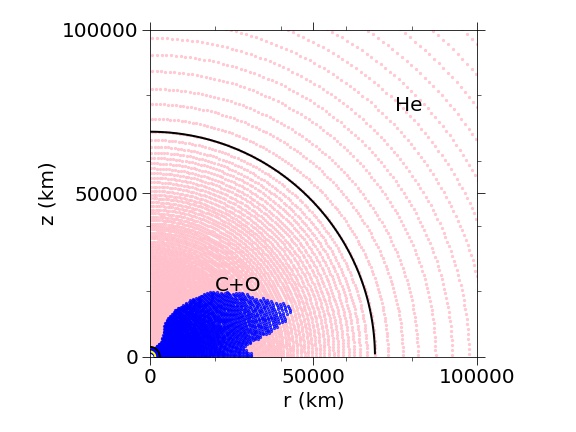}
    \includegraphics[width=0.48\textwidth]{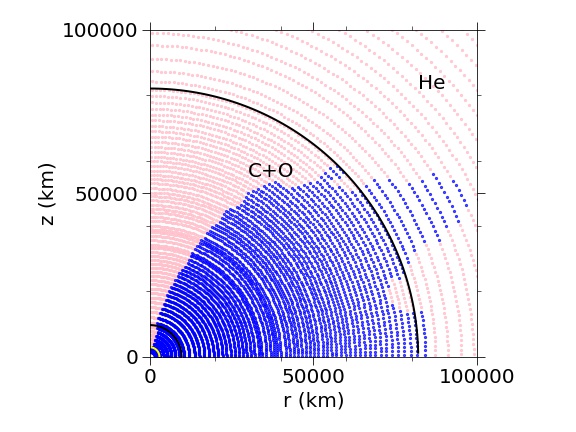}
    
    \caption{The initial tracer distribution for both bound by gravity (blue circles) and ejected (pink triangles). The annotations and circles correspond to the element shell from the pre-explosion models for M20-100-100 (top panel), M30-100-100 (middle panel) and M40-100-100 (bottom panel).}
    \label{fig:mass_tracer_plot}
\end{figure}

In Figure \ref{fig:mass_tracer_plot} we plot the distribution of ejecta and bound tracers accordingly to their final total energy. The geometry where the tracers are ejected is closely related to the progenitor ZAMS mass. We choose each of the progenitors with the jet energetics from the characteristic model, i.e., MXX-100-100 with XX = 20, 30, and 40. For the lower mass model, only the zone directly normal to the jet direction up to 20000 km experiences later accretion. With the majority of stars being ejected, the explosion is quasi-spherical. However, we remarked that the tracer thermodynamics still experience angle-dependent variation. 

For the $30M_{\odot}$ model, the zone extends broadly from the Si to the middle of the C+O core, with an effective open angle of $\sim 30^{\circ}$. Tracers near the axis are more readily ejected than those along the diagonal direction. The explosion appears more aspherical than we expect, with more Si-group elements along the pole and less so along the equator. 

For the $40M_{\odot}$ model, almost the entire Si core and a majority of the C+O core remain bound after the jet energy deposition. Within the Si core, the ejected zone agrees with the open angle of the jet. In the C+O core, the affected part expands again to $\sim 30^{\circ}$ all the way to the He core. Part of the lower He-core remains bound after the explosion. This is expected because of the much higher C+O core mass. The explosion is much more aspherical -- near the jet-axis, Fe-group and Si-group elements are ejected while the ejecta along equators are more He-rich. 

\begin{figure*}
    \centering
    \includegraphics[width=0.95\textwidth]{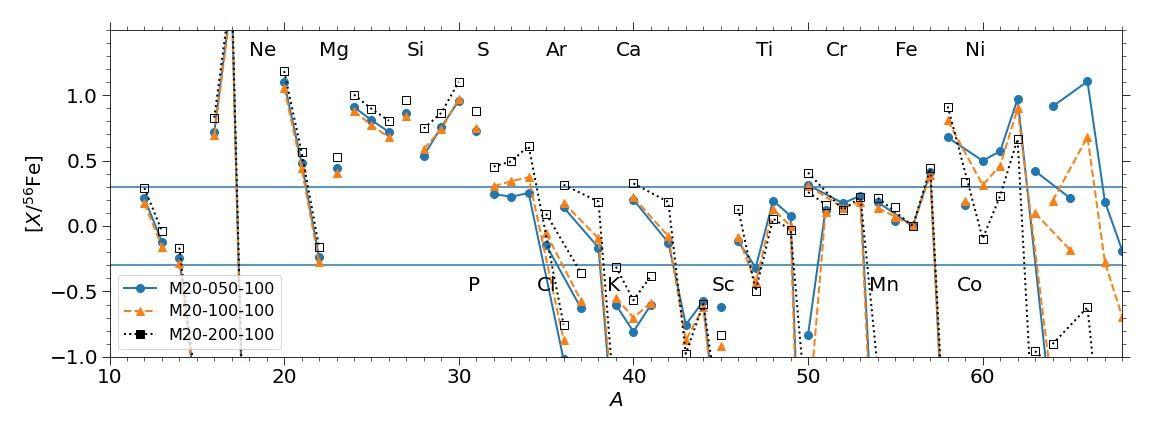}
    \includegraphics[width=0.95\textwidth]{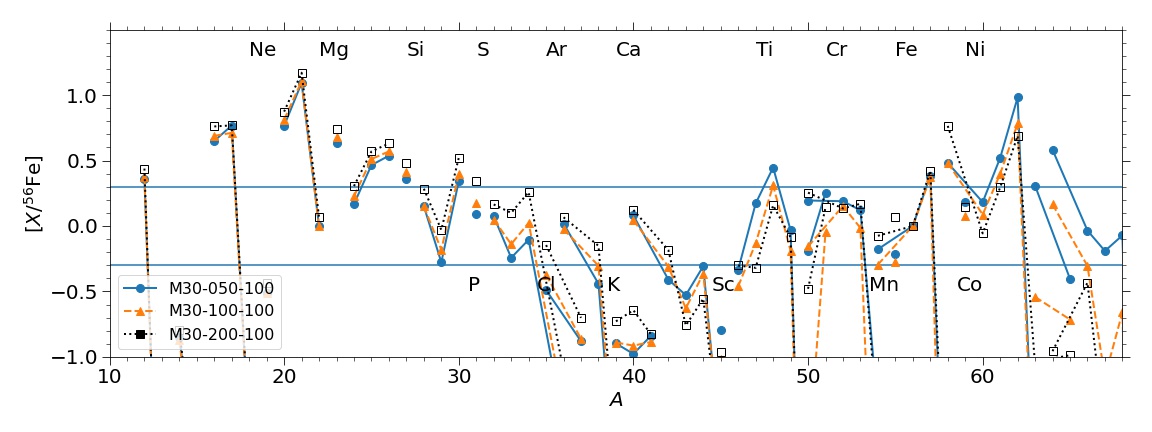}
    \includegraphics[width=0.95\textwidth]{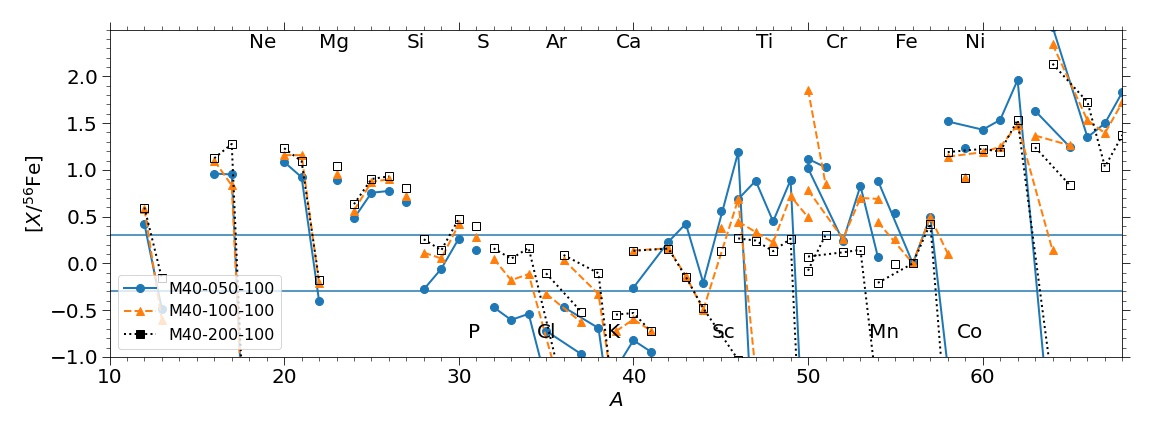}
    
    \caption{(top panel) The isotopic mass fraction [X/$^{56}$Fe] (in solar unit) for M20-050-100, M20-100-100 and M20-200-100.
    (middle panel) Same as the top panel but for M30-050-100, M30-100-100 and M30-200-100.
    (bottom panel) Same as the top panel but for M40-050-100, M40-100-100 and M40-200-100.}
    \label{fig:mass_xiso_plot}
\end{figure*}

In Figure \ref{fig:mass_xiso_plot} we plot how the chemical composition depends on the energy deposition rate to illustrate how the explosive nucleosynthesis is affected by a weak, medium, or strong jet. In the top panel, we include M20-050-100, M20-100-100, and M20-200-100. They share the same jet deposition duration but different deposition rates. Due to the almost complete disruption, the chemical yields are very similar among models. Only for M200-200-100, we see a slight enhancement in the ratios compared to the other two models: Isotopes before S are overproduced; isotopes from S to Fe are well produced, including Co and Mn, but K and Sc are underproduced. The most significant difference appears in Zn, where a strong jet suppresses the production of Zn.  

In the middle panel, we show a similar plot but for M30-050-100, M30-100-100, and M30-200-100. We also see highly similar nucleosynthetic yields. Only for Fe-group elements like $^{55}$Mn, $^{58}$Ni $^{63,65}$Cu, and ${64}$Zn do we find differences. This reflects that in this mass range, the jet only changes the shock strength in the Si-core, while in the O-core, the shock loses energy and does not vary much across models.

In the bottom panel, we also plot for M40-050-100, M40-100-100, and M40-200-100. The impact of shock becomes the most pronounced because it affects both the shock strength and the mass ejection, which change the relative contribution from each shell. A stronger jet leads to a higher fraction of Si-group elements, including Si, P, S, and Cl. This is related to the ejected region of the C+O core. Meanwhile, the lower Fe-group elements up to Fe have a flipped dependence: a strong jet reduces the mass fraction of those isotopes. It is because a stronger jet synthesizes more $^{56}$Fe in the same explosion. The mass fraction ratio faces a larger denominator. Lastly, for upper Fe-group elements, the production is counteracted by two factors: stronger jet supports production of these isotopes, and the higher $^{56}$Fe balances the ratios. 

\begin{figure}
    \centering
    \includegraphics[width=0.48\textwidth]{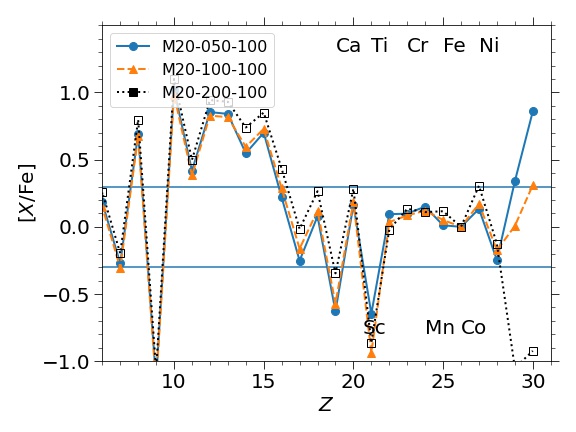}
    \includegraphics[width=0.48\textwidth]{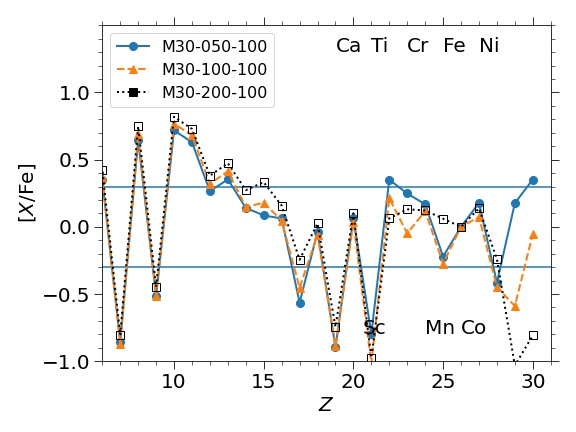}
    \includegraphics[width=0.48\textwidth]{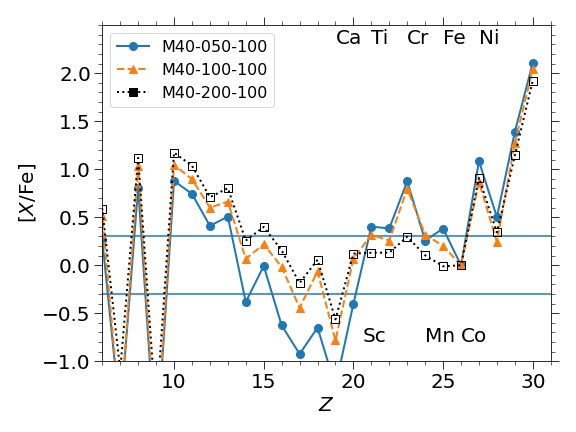}
    
    \caption{(top panel) The elemenal mass fraction [X/$^{56}$Fe] (in solar unit) for M20-050-100, M20-100-100 and M20-200-100.
    (middle panel) Same as the top panel but for M30-050-100, M30-100-100 and M30-200-100.
    (bottom panel) Same as the top panel but for M40-050-100, M40-100-100 and M40-200-100.}
    \label{fig:mass_xele_plot}
\end{figure}

In most spectroscopic observation, the spectra do not distinguish among isotopes. Thus in Figure \ref{fig:mass_xele_plot} we also plot the elemental abundance of the same three sets of models. 

In the top panel, we show the M20 series for the weak, medium, and strong jets. The almost identical abundance pattern reconfirms that the explosion disrupts the entire star and hence all elements are similarly produced. A weaker jet leads to a much higher Zn production. This could be understood by the thermodynamics history that for a very strong jet, while the temperature could be high enough for the nuclear reaction to proceed along the $\alpha$-chain, the duration could be shorter than that in a weak jet. 

In the middle panel, we plot the models in the M30 series. Most elements up to Sc are similar. Some odd-number elements, such as V and Mn, show non-monotonic behaviour. A stronger jet by comparing M30-050-100 and M30-100-100 gives a lower yield of these elements, but an even stronger jet by comparing M30-100-100 and M30-200-100, gives a higher yield. This is also related to the competition between Fe production and the electron capture process, which happens for high-density matter in the nuclear statistical equilibrium. 

In the bottom panel, we plot the elements for the M40 series. Agreeing with the isotopic distribution, a stronger jet results in more Si-group elements due to higher mass ejection, but a lower Fe-group elements from Sc to Fe due to the higher Fe production. Elements beyond Fe are similar among all three models. 

\section{Applications}

The extreme metal-poor stars (EMPSs) are most likely enriched by a single or a few massive star explosions in the early universe \citep[e.g., ][]{Hartwig2023EMP}. The elemental abundance of these stars thus provides a robust window to compare individual supernova explosion models with less concern about the cumulative effects from generations of other supernovae in the supernova history. In \cite{Jeong2023EMPS}, 18 extremely metal-poor (EMP; [Fe/H]$<–3.0$) stars surveyed by the Sloan Digital Sky Survey (SDSS) and Large Sky Area Multi-Object Fiber Spectroscopic Telescope (LAMOST) survey are measured by the high-resolution telescope using GEMINI-N/GRACES. These stars show signatures compatible with typical halo stars. Here we compare the new collapsar models with these stellar yields. 

\begin{figure}
    \centering
    \includegraphics[width=0.95\linewidth]{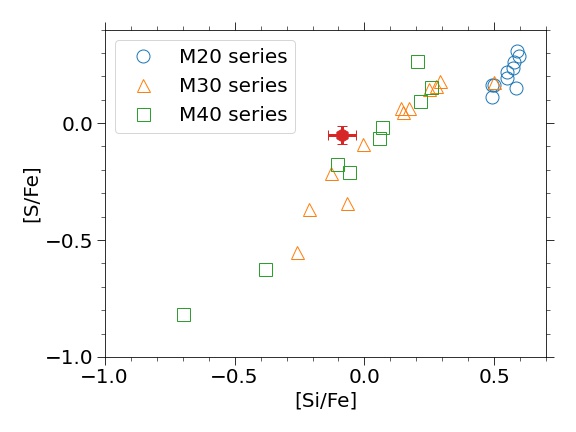}
    \includegraphics[width=0.95\linewidth]{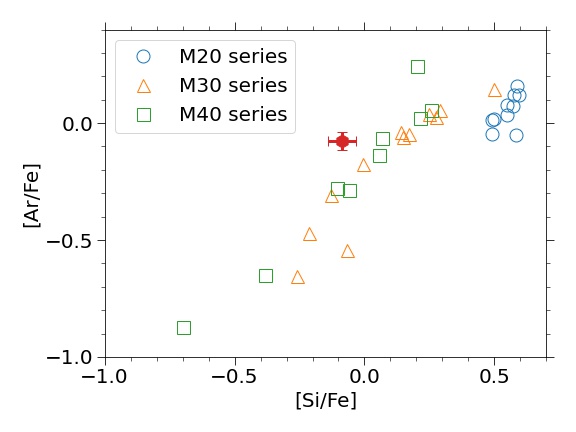}
    \includegraphics[width=0.95\linewidth]{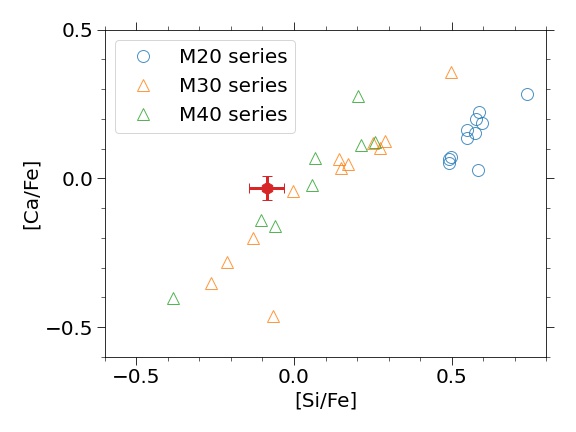}
    \caption{(top panel) The [S/Fe] against [Si/Fe] for the models presented in this work, with observational data obtained from the Perseus Cluster. 
    (middle panel) Same as the top panel, but for [Ar/Fe] against [S/Fe].
    (bottom panel) Same as the top panel, but for [Ca/Fe] against [S/Fe].}
    \label{fig:perseus_plot}
\end{figure}

In the upper panel of Figure \ref{fig:perseus_plot}, we plot the [S/Fe] against [Si/Fe] of the jet-induced explosion models presented in this work, together with the data taken from the Perseus Cluster \citep{Simionescu2019Perseus}. All models appear with a similar slope $\sim 1$, because the formation channels of Si and S are extremely similar. The M20 series clusters at higher [Si/Fe] and [S/Fe] values with smaller dispersions. The M30 and M40 have similar spreads and values. The near solar ratio of the Perseus Cluster is very close to the M30 and M40 series, but is higher in both element ratios. 

In the middle panel, we plot similar to the top panel but with the y-axis replaced by [Ar/Fe]. The cluster within the series and the spreads are very similar to [S/Fe]. The Perseus Cluster also shows a value pair close to the solar values. The M30 and M40 series are closest to the data point. The underproduction of [Ar/Fe] becomes more significant, which differs by $\sim0.1-0.2$ dex. 

In the bottom panel we plot similar to the top panel but for [Ca/Fe]. The distinction between M20 and the other two groups the other two groups are also clear. Based on the three elements, it suggests the possibility of identifying the dominant mass based on the typical range of the elements in the SN ejecta. The proximity of the Perseus Cluster with the 30--40 $M_{\odot}$ could imply an inverted population of collapsar, opposite to the Salpaeter relation for ordinary stars. 

\begin{figure}
    \centering
    \includegraphics[width=0.95\linewidth]{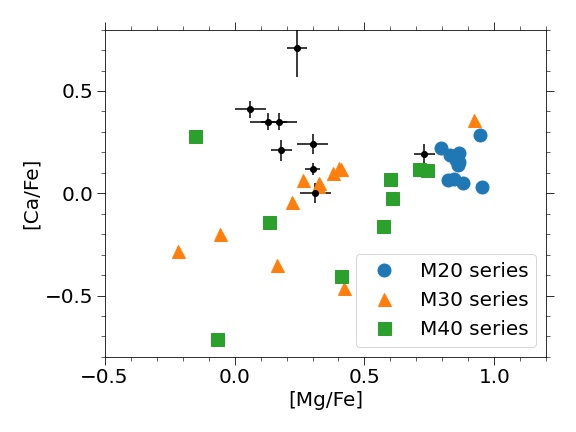}
    \includegraphics[width=0.95\linewidth]{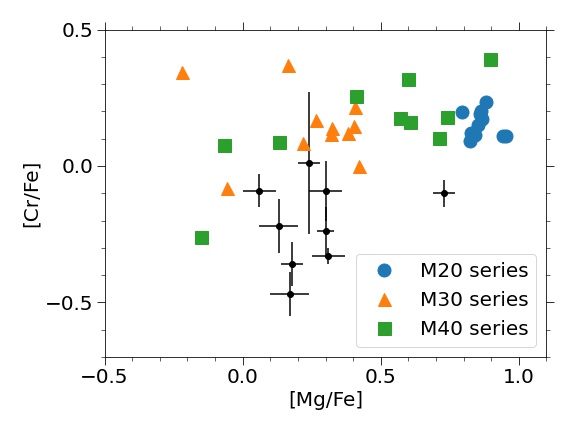}
    \caption{(top panel) The [Ca/Fe] against [mg/Fe] for the models presented in this work, with observational data from metal poor stars. 
    (bottom panel) Same as the top panel, but for [Cr/Fe] against [S/Fe].}
    \label{fig:emps_plot}
\end{figure}

Then we apply the collapsar models to the extremely metal-poor stars reported in \cite{Jeong2023EMPS}. Metal-poor stars are expected to be chemically enriched by one or a few supernovae only. Thus, these chemical abundances could provide extra insight into the individual supernova explosion models. In Figure \ref{fig:emps_plot}, we plot the [Ca/Fe] against [Mg/Fe] for our models. The data points are strongly enriched in Ca, where [Ca/Fe] $\sim 0.0 - 0.8$. The [Mg/Fe] is typically near solar, with one exception of about 0.7. The models can generate the spread in [Mg/Fe], but not [Ca/Fe]. Most models predict only near solar or sub-solar [Ca/Fe]. Such a high value typically requires a more exotic explosion channel, such as a Ca-rich or the lower end of massive star explosion (e.g., electron-capture supernovae). 

In the bottom panel, we plot the [Cr/Fe] against [Mg/Fe]. The data point occupies the sub-solar values of [Cr/Fe]. In contrast, the collapsar models show above-solar values for most models. The high Cr/Fe ratio in the collapsar model is expected because of the high entropy environment inside the bipolar cone-shaped zone. Such a high entropy, although the density is not high enough for nuclear statistical equilibrium, could push the nuclear reactions along the $\alpha$-chain to the lower-half of iron-group elements (Ti, V, Cr). 

\begin{figure}
    \centering
    \includegraphics[width=0.9\linewidth]{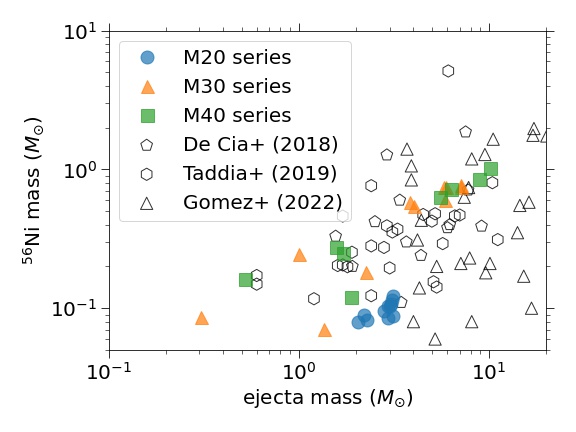}
    \caption{The $^{56}$Ni yield against ejecta for the models presented in this work. Observational data are taken from supernova surveys for superluminous SNe Ib/c \citep{DeCia2018}, SNe Ic-BL \citep{Taddia2019}, and luminous SNe Ib/c \citep{Gomez2022}. }
    \label{fig:56Ni_Mej_plot}
\end{figure}

In Figure \ref{fig:56Ni_Mej_plot} we plot the $^{56}$Ni against the ejecta mass $M_{\rm ej}$ for the collapsar model. We also add the $^{56}$Ni and $M_{\rm ej}$ from the intermediate Palomar Transient Factory (iPTF) \citep{Law2019PTF}.
We use the catalogues documented for superluminous SNe Ib/c \citep{DeCia2018}, SNe Ic-BL \citep{Taddia2019}, and luminous SNe Ib/c \citep{Gomez2022} to expand the diversity of the samples. The supernova samples show an approximately uniform distribution for $M_{\rm Ni}$ between (0.05, 2) $M_{\odot}$ and $M_{\rm ej}$ between (1, 20) $M_{\odot}$. The models presented in this work occupy the lower right region of the figure. The models can better explain the spread reported in \cite{Taddia2019}, but not \cite{Gomez2022}. 

The models show a positive correlation in the observable pairs, which is expected as higher explosion energy implies more matter undergoing complete burning. This also implies that the jet-driven supernova models cannot explain the lower right region, which corresponds to high ejecta mass but low $^{56}$Ni production.

\begin{figure}
    \centering
    \includegraphics[width=0.9\linewidth]{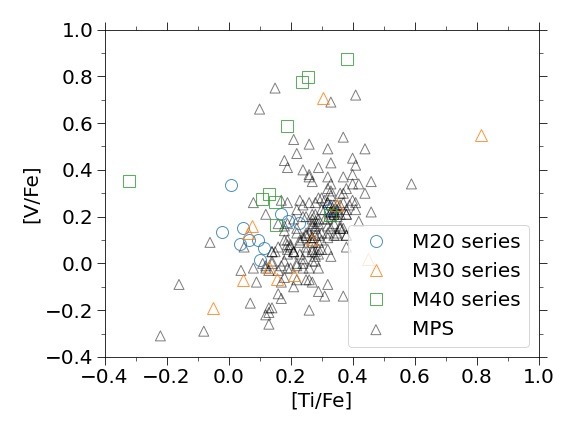}
    \caption{The [V/Fe] against [Ti/Fe] for the models presented in this work, with observational data from metal-poor star survey reported in \citep{Roederer2014}. Only stars with determined values of [Ti/Fe] and [V/Fe] are plotted.}
    \label{fig:VFe_TiFe_plot}
\end{figure}

We also compare the element pair [V/Fe] against [Ti/Fe] for our jet-driven supernova models with the observed metal-poor star reported in the catalogue in the HK-II survey \citep{Roederer2012,Roederer2014}. The neutron numbers in $^{50}$V and $^{51}$V are high, which are representative isotopes for matter having experienced electron capture. Ti is the general product of incomplete burning along the $\alpha$-chain. These elements pair probe the two aspects of the explosion. The data point shows a clustered group around [V/Fe] between (-0.1, 0.3) and [Ti/Fe] between (0.0, 0.4), with a positive correlation between the two element ratios. Our models also show a similar range that overlays the observed cluster. 

In \cite{Leung2024Jet2}, we have performed a similar survey using the pre-supernova models presented in \cite{Nomoto2013CCSNYield}. The jet-driven supernova models presented in that one show the models sharing a similar slope as the observational data, but the overall intercept is higher, thus the model sequence is visually off from the cluster. The models presented in this work show both the slope and intercept much closer to the cluster. This provides further confidence that the observed diversity in these metal-poor stars could be connected to some past explosion history of jet-driven supernovae. 

\section{Role of Collapsar in the Galactic Scale}
\label{sec:GCE}

\begin{table*}[]
    \centering
    \caption{The models for the galactic chemical evolution. The columns ``CCSN'', ``SN Ia'' and ``Collapsar'' stands for the massive star, Type Ia supernova explosion models used. $D_{\rm col}$ and $Z_{\rm off}$ represent the coefficient for collapsar (see Eq. \ref{eq:GCE}). The $\chi^2$ corresponds to the total $\chi^2$ in each of the parameter survey by comparing with Milky Way stars from the SAGA database \citep{SAGA2008}. The column ``population'' refers to if the number fraction of the 20, 30, and 40 $M_{\odot}$ stars are determined by the Salpaeter relation or the value of $f_{\rm top}$ by the parametrized model. The [Si/Fe], [S/Fe], [Ar/Fe] and [Ca/Fe] are the scaled mass fraction of that element for the best model, in solar unit defined as 
    [X/Fe]$= \log_{10}[$($X$/Fe)/($X$/Fe)$_{\odot}$].}
    \begin{tabular}{c c c c c c c c c c c c }
        \hline
         Model & CCSN & SN Ia & Collapsar & $D_{\rm col}$ & $Z_{\rm off}$ & Population & $\chi^{2}$  & [Si/Fe] & [S/Fe] & [Ar/Fe] & [Ca/Fe] \\ \hline
         LN25-LN18Ka4-S & LN25 & LN18(Ka4) & Strong & 0.20 & 0.012 & Salpaeter & 54.38 & -0.027 & -0.081 & -0.116 & -0.099\\
         LN25-LN18Ka4-M & LN25 & LN18(Ka4) & Med & 0.20 & 0.01175 & Salpaeter & 63.91 & -0.027 & -0.081 & -0.116 & -0.099\\
         LN25-LN18Ka4-W & LN25 & LN18(Ka4) & Weak & 0.31 & 0.01175 & Salpaeter & 66.76 & -0.027 & -0.081 & -0.116 & -0.099 \\
         LN25-SK18-S & LN25 & SK18 & Strong & 0.20 & 0.0115 & Salpaeter & 54.57 & 0.010 & -0.053 & -0.083 & -0.094  \\
         LN25-SK18-M & LN25 & SK18 & Med & 0.20 & 0.01125 & Salpaeter & 63.74 & 0.010 & -0.053 & -0.083 & -0.094 \\
         LN25-SK18-W & LN25 & SK18 & Weak & 0.31 & 0.01125 & Salpaeter & 66.46 &  0.010 & -0.053 & -0.083 & -0.094 \\ \hline
         LN25-LN18Ka4-Sb & LN25 & LN18(Ka4) & Strong & 0.30 & 0.025 & 0.16 & 116.20 & -0.027 & -0.081 & -0.116 & -0.099  \\ %-0.128 \\
         LN25-LN18Ka4-Mb & LN25 & LN18(Ka4) & Med & 0.19 & 0.025 & 0.26 & 138.43 & -0.027 & -0.081 & -0.116 & -0.099 \\
         LN25-LN18Ka4-Wb & LN25 & LN18(Ka4) & Weak & 0.11 & 0.025 & 0.8 & 146.58 & -0.027 & -0.081 & -0.116 & -0.099\\
         LN25-SK18-Sb & LN25 & SK18 & Strong & 0.35 & 0.025 & 0.13 & 121.32 & 0.010 & -0.053 & -0.083 & -0.094 \\
         LN25-SK18-Mb & LN25 & SK18 & Med & 0.26 & 0.025 & 0.18 & 152.27 & 0.010 & -0.053 & -0.083 & -0.094 \\
         LN25-SK18-Wb & LN25 & SK18 & Weak & 0.16 & 0.025 & 0.51 & 163.96 & 0.010 & -0.053 & -0.083 & -0.094 \\ \hline
    \end{tabular}
    \label{tab:best_model_Perseus}
\end{table*}

\subsection{Modified Equations}

The GCE code solves the time evolution of individual isotopes by including their production (stellar and supernova nucleosynthesis, inflow) and their destruction (star formation, stellar outflow). In this work, we extend the calculation by including the collapsar yield as a new source 
\begin{equation}
\begin{aligned}
\frac{d\sigma_i}{dt} &= 
\int_{0.8}^{3} B(t - \tau(m)) \, \Psi(m) \times X_i(t - \tau(m)) \, dm  \\
&\quad + C \int_{3}^{16} \int_{\mu_m}^{0.5} B(t - \tau(m)) \, \Psi(m) \, f(\mu) \\
&\quad \times X_i(t - \tau(m)) \, d\mu \, dm \\
&\quad + (1 - C) \int_{3}^{16} B(t - \tau(m)) \, \Psi(m)  \\
&\quad \times X_i(t - \tau(m)) \, dm \\
&\quad + (1 - D_{\rm Col}) \int_{16}^{40} B(t - \tau(m)) \, \Psi(m) \times X_i(t - \tau(m)) \, dm \\
&\quad + D_{\rm Col} \int_{16}^{40} B(t - \tau(m)) \, \Psi(m) \times  X_i(t - \tau(m)) \, dm \\
&\quad - B(t) \frac{\sigma_i}{\sigma_{\rm gas}} + \dot{\sigma}_{i,\rm gas} + \frac{\sigma_i}{\tau_{1/2}} M_\odot \, \text{pc}^{-2} \, \text{Gyr}^{-1}
\end{aligned}
\label{eq:GCE}
\end{equation}

As a remark, $\frac{d\sigma_i}{dt}$ is representative of the rate of change of the surface mass density of an isotope $i$, including relevant terms to stellar death, birth, gas infall, and isotope decay \citep{Timmes1995GCE}. $B(t)$ represents the birthrate function at time $t$. $\tau(m)$ accounts for different lifetimes of main-sequence stars at mass $m$ and $\Psi(m)$ is the initial mass function. The term $X_i(t - \tau(m))$ represents the abundance of element $i$ in stars formed at time $t - \tau(m)$. The free parameter $C$ is the fraction of intermediate-mass stars (3 -- 8 $M_{\odot}$) in binary systems that produce Type Ia supernovae, with $f(\mu)$ describing the distribution of binary mass ratios $\mu$. 
%In the last line, $B(t) \frac{\sigma_i}{\sigma_{\rm gas}}$ is the current star formation rate times the proportion of $i$ in the total gas. $\dot{\sigma}_{i,\rm gas}$ is the infall or outflow rate of isotope $i$. $\frac{\sigma_i}{\tau_{1/2}}$ serves as the radioactive decay. 
The last line contains the gas inflow, outflow, and radioactivity source terms. 
The second-last line is the newest addition to our collapsar explosion channel. The new parameter $D_{\rm Col}$ is the fraction of massive stars that undergo aspherical collapsar CCSN. 

We assume that a certain fraction of massive stars explode as jet-driven supernovae, controlled by the parameter $D_{\rm col}$. In general, this rate could be dependent on two factors: (1) the statistics of the angular rotation of stars, and (2) the critical metallicity where the mass loss might suppress the formation of a rapidly-rotating black hole. In this work, we take it as a constant ($D_{\rm col} \in (0,1)$ so that we can quantify the contribution of these collapsars.

In searching for the best models, besides $D_{\rm col}$, we also introduce the parameter $Z_{\rm off}$, which corresponds to the metallicity above which the collapsars stop forming. This happens for high metallicity stars, where the enhanced mass loss during the H- and He- main-sequence removes most of the angular momentum in the outer envelopes. The resultant Fe-core and the infalling matter will be slowly rotating, which fails to induce the spontaneous energy outburst. In this work, we also treat this parameter as a constant with $Z_{\rm off} \in (0,0.025)$ in the parameter surveys. 

We also consider in some models the possibility of a stellar population different from the Salpeter relation. This relation is calibrated based on the Milky Way stars, which correspond to the current universe. The number fraction of the 20, 30, and 40 $M_{\odot}$ stars could be different in early galactic systems, where massive stars formed first and have shorter lifetimes. In view of that, we introduce the parameter $f_{\rm top}$ which parametrizes the mass fractions of these three groups of stars as $0.67 (1-f_{\rm top}), 0.33, 0.67 f_{\rm top}$. This covers both the bottom-heavy (Salpaeter), mass independent ($f_{\rm top} = 0.5$), and top-heavy $f_{\rm top} > 0.5$ cases. 

\subsection{Application to the Milky Way}

In Table \ref{tab:best_model_Perseus} we list the parameter surveys run in this article. It assumes the same massive star models as in this series of articles. We consider two SNe Ia models, including SK18 \citep[S1.0, ][]{Shen2018DDDDDD} and LN25Ka4 \citep[300-1-c3-1, with Karlovitz number 4.0][]{Leung2018Chand} to explore the effect of the Ch-mass and subCh-mass WDs. For the collapsar models, we use the models with suffix 050-100 for the ``weak'' model, 100-100 for the ``medium'' model and 200-100 for the ``strong'' model. They correspond to the different energy deposition strengths by the collapsar. In Table \ref{tab:models}, we show that they have a strong impact on the ejecta mass and the overall composition of the ejecta. Each series consists of massive star models of 20, 30, and 40 $M_{\odot}$. Their relative fraction follows the Salpeter relation $dn(m) \sim m^{\alpha}$ where $\alpha \approx -2.35$.  

To compute the $\chi^2$-fitting for the chemical mass fraction ratios with the stars, we collect the abundance data from the SAGA database and bin them by their metallicity at a 0.5 interval between (-3, 0), and denote their average metallicity $Z_i$, the average abundance ratio $X_{{\rm SAGA}, i}$ and its standard deviation $\sigma_{{\rm SAGA},i}$. This gives us a total of 6 data points. Then we compare the abundance ratio at the same average metallicity in the GCE model, by 
\begin{equation}
    \chi^2 = \sum_{i=1}^6 \frac{(X_{\rm GCE}(Z_i) - X_{{\rm SAGA}, i})^2}{\sigma_{{\rm SAGA}, i}}.
\end{equation}
The total $\chi^2$ is computed by the $\chi^2$ of these reference elements. 

%\red{Henry: Please add the figure for chi-sq colour plots and add the discussion similar to what we do for the Perseus2}

\begin{figure*}
    \centering
    \includegraphics[width=0.48 \textwidth]{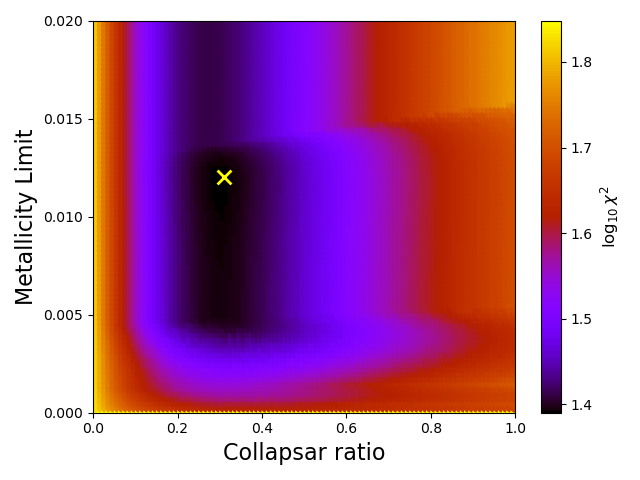}
    \includegraphics[width=0.48 \textwidth]{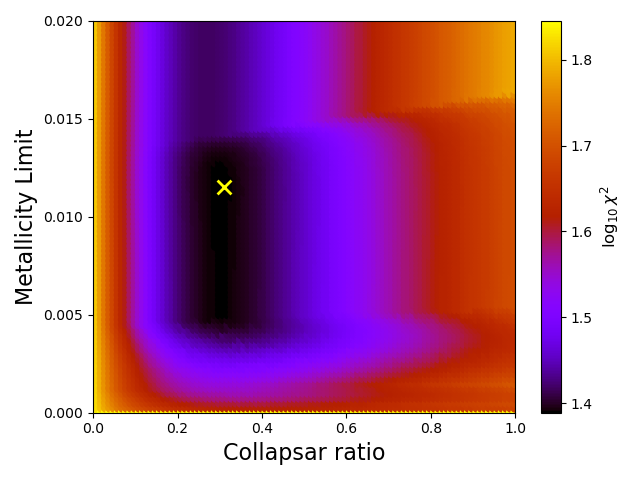}
    \includegraphics[width=0.48 \textwidth]{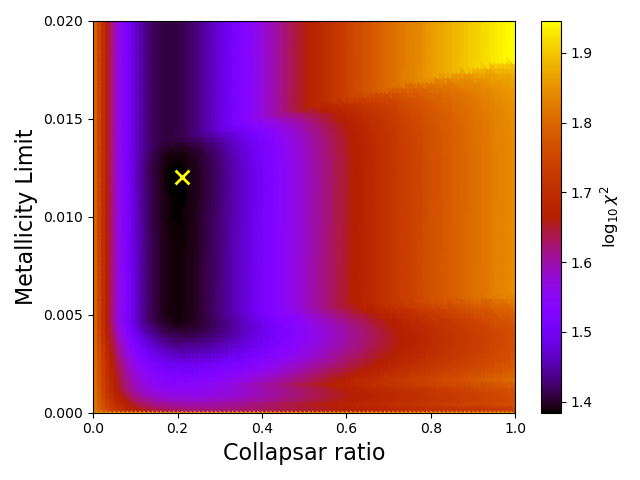}
    \includegraphics[width=0.48 \textwidth]{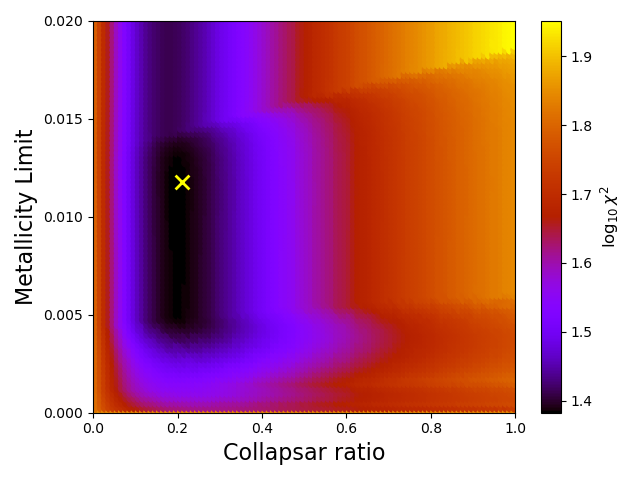}
    \includegraphics[width=0.48 \textwidth]{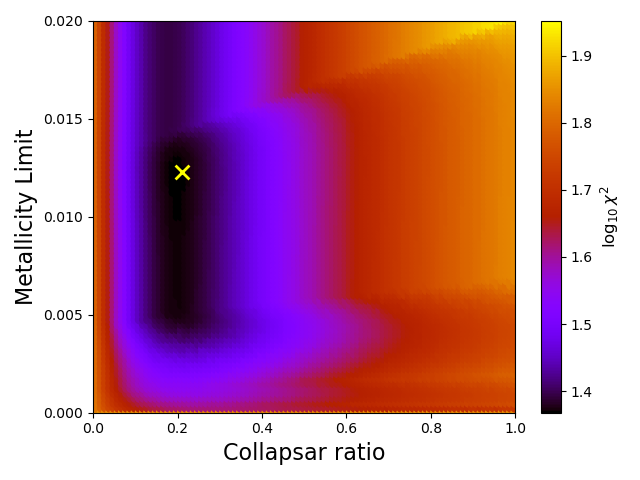}
    \includegraphics[width=0.48 \textwidth]{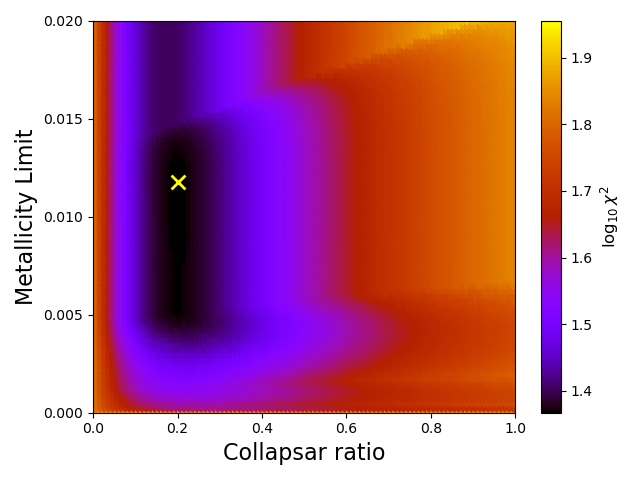}
    \caption{The left column shows the best runs for models including the LN18(Ka4) SN Ia yields, given data from the Milky Way via the SAGA database; the right column shows the same for models including the SK18 SN Ia yields. The top row shows the runs from models using the weak collapsar yields, the second row shows runs corresponding to the medium collapsar yields, and the last row shows the runs corresponding to the strong collapsar yields. The yellow cross represents each models lowest chi-square value from various models with varying $Z_{\rm off}$ and $D_{\rm col}$. }
    
    \label{fig:GCE_chisq}
\end{figure*}

We perform the parameter survey for the six supernova yield sets by using $D_{\rm col}$ and $Z_{\rm off}$ as the model parameters. In Figure \ref{fig:GCE_chisq} we show the colour plot for the $\chi^2$-fitting of these surveys. This explores both the effects of SNe Ia and collapsars. We include SNe Ia because, from Paper II, the SNe Ia have an influential change to the interpretation of the SN rates. 

The six figures look very similar to each other, where the best models are marked by the yellow cross. A high collapsar ratio is not favoured. This also agrees with the rarity of the collapsar or GRB events relative to the SN rates. A fraction about $\sim 0.2 - 0.3$ are preferred. At the same time, all models show a similar conclusion that collapsar should have no contribution beyond $Z \geq 0.012$. The similarity of our fitting with different SNe Ia models and collapsar energetic confirms that these constraints apply to collapsar in general. 

%\red{Task 13: Then please add the evolution trends for the best models for all 6 choices.}

\begin{figure*}
    \centering
    \includegraphics[width=0.48\linewidth]{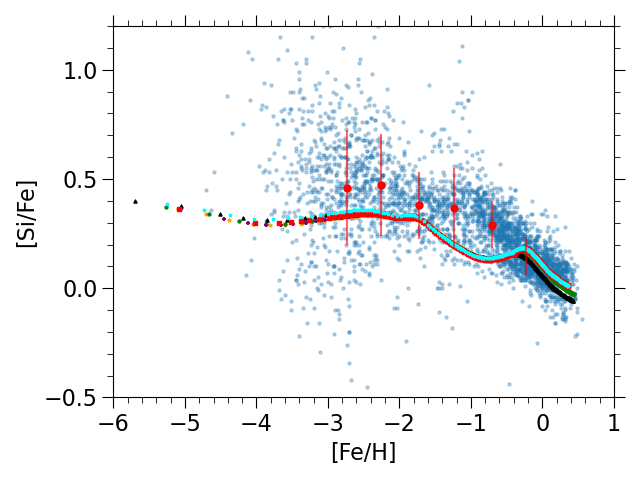}
    \includegraphics[width=0.48\linewidth]{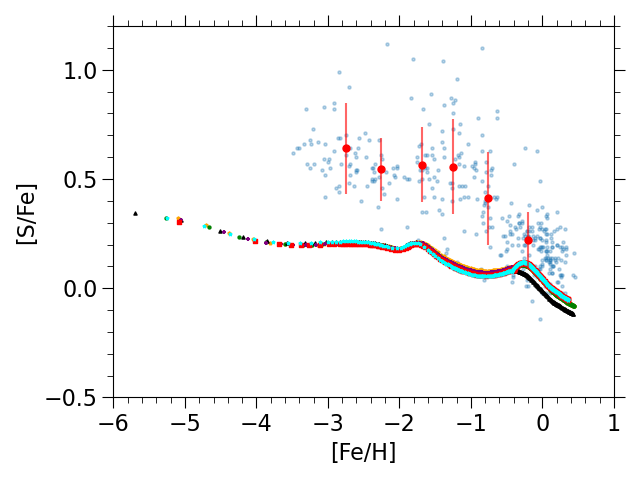}
    \includegraphics[width=0.48\linewidth]{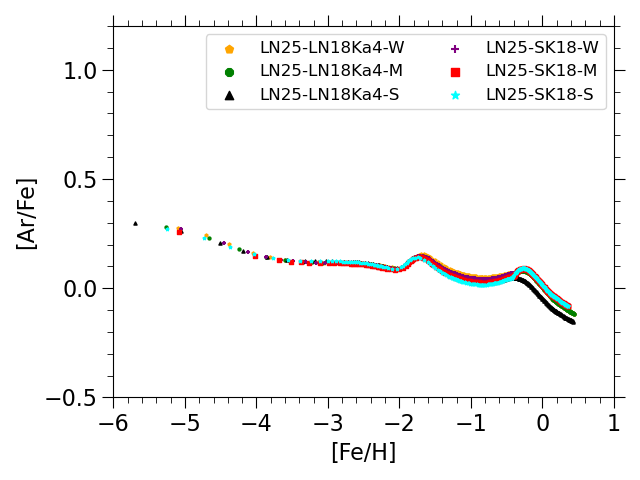}
    \includegraphics[width=0.48\linewidth]{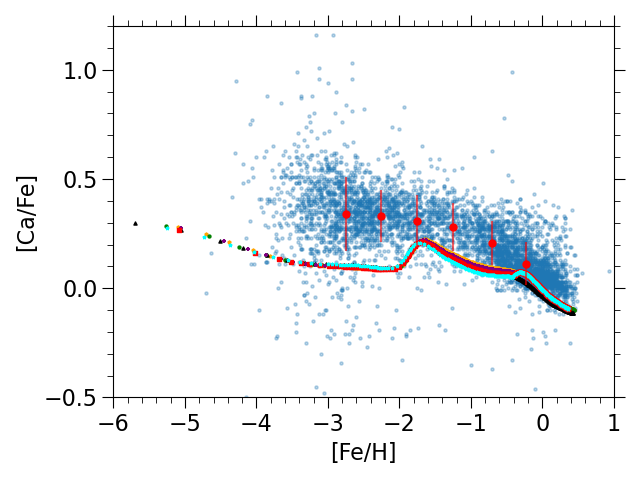}
    \includegraphics[width=0.48\linewidth]{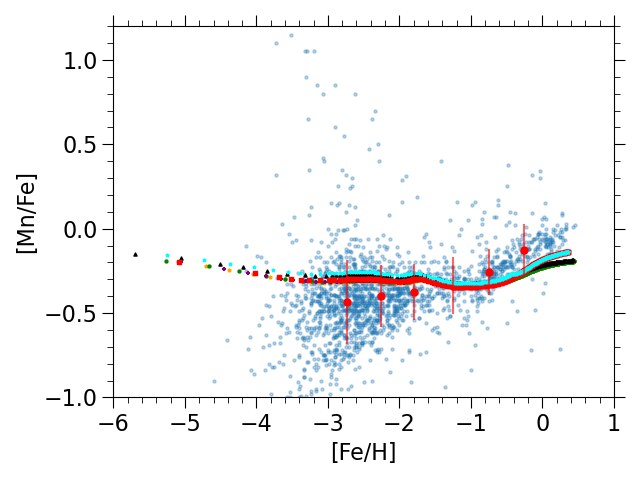}
    \includegraphics[width=0.48\linewidth]{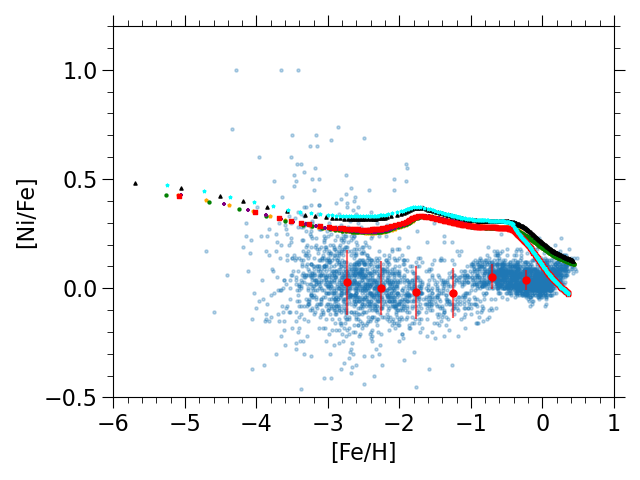}
    \caption{The chemical element evolution for the best models from Table \ref{tab:best_model_Perseus}. The scattering points are the stellar abundances for stars in the Milky Way \citep{SAGA2008}, including both main-sequence and red-giant branch stars. The error bars are the metallicity-binned averages and standard deviations derived from the stellar abundances. }
    \label{fig:xfe_z_plot}
\end{figure*}

In Figure \ref{fig:xfe_z_plot}, we plot the element mass ratio [X/Fe] as a function of the local metallicity. The best-fit models are selected based on the mass fraction of selected elements at given metallicity, indicated by the metallicity bins from the stellar data. 

For Si, the best models can mimic mostly the falling trend of the data, but the lines lie at the lower side of the data. The collapsar population creates a second knee identified at [Fe/H] $= -2$, and the SN Ia population creates the second knee at -0.5. 

A similar knee pattern is observed for [S/Fe] and [Ca/Fe]. Similar to Paper II, the high S/Fe indicates additional astrophysical sources not identified in current massive star or collapsar models. Although the trends in the best models follow those in the stellar data. Similar observations can be obtained for [Ca/Fe].

For Ar, there is no data from the SAGA database. All collapsar models show very similar trends and deviate by less than 0.1 dex near solar metallicity. 

For Mn, the new massive star and collapsar models lead to a higher Mn/Fe at low metallicity [Fe/H] $< -2$. This again confirms the importance of the massive star models in reconciling the Mn/Fe production on the galactic scale. The massive star models allow the parameter survey to pick a smaller proportion of SNe Ia to the rising trend at [Fe/H] $= -1$. The collapsar makes no observable difference in the Mn trend. 

For Ni, the details of the collapsar also do not play an active role in the trend. The massive star is responsible for the magnitude up to [Fe/H] $< -0.5$, which is 0.3 dex too high from the average of the star. There is a slight increase in [Ni/Fe] for the stellar data around [Fe/H]$\sim -1$. The first knee could be related to this collapsar transition. However, the stellar data do not show a strong drop in [Ni/Fe] near solar metallicity. The drop is likely a consequence of the $\chi^2$ fitting where models with near-solar Ni/Fe are preferred. This forces the code to search for models where Ni/Fe is diluted by later SN explosions. 

\subsection{Parameter Dependences of Collapsar on Zn production}

The collapsar is known to produce Zn-rich material because the focused energy deposition creates a high entropy environment. Some hyper metal-poor stars (HMPS, [Fe/H] $< -5$) such as HE 1327-2326 have been measured with a super-solar value of [Zn/Fe] = $0.80 \pm 0.25$. Note that Zn/Fe in typical massive star models is always near solar value. The discovery of this HMPS indicates that the collapsar channel is highly active in the early universe. 

\begin{figure}
    \centering
    \includegraphics[width=0.98\linewidth]{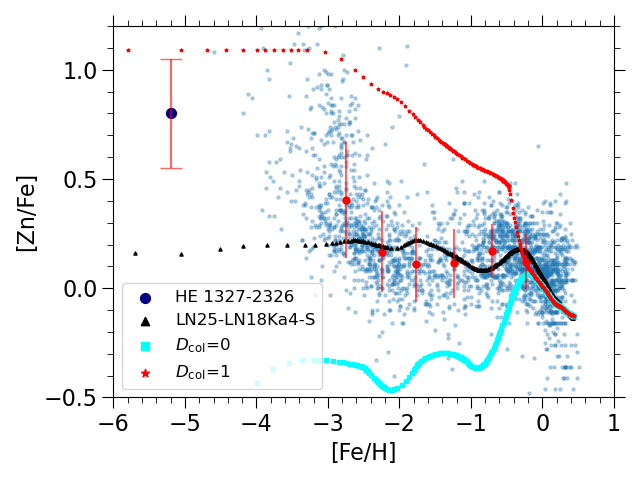}
    \includegraphics[width=0.98\linewidth]{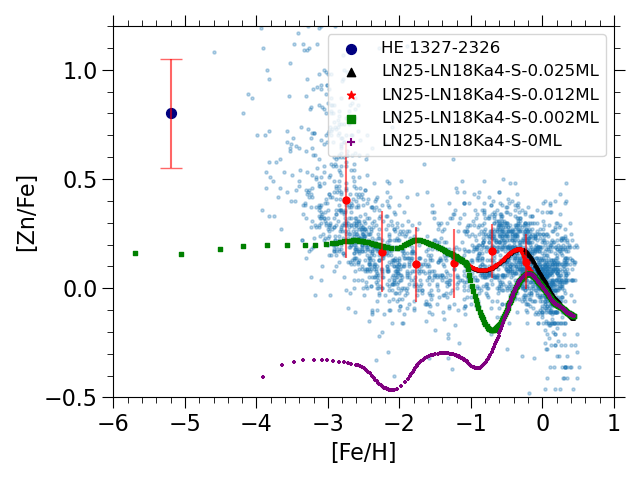}
    \includegraphics[width=0.98\linewidth]{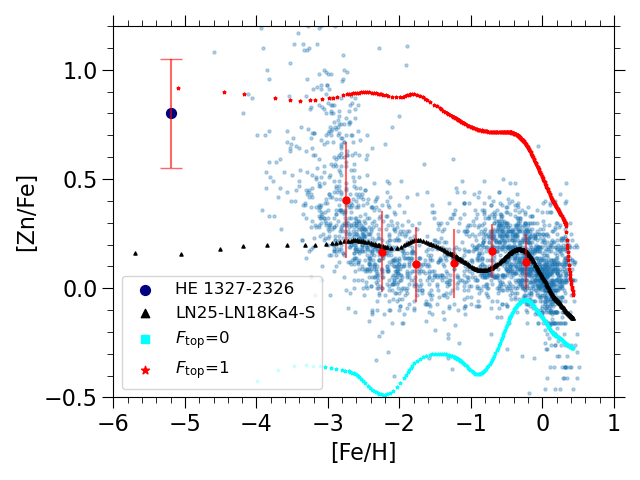}
    \caption{(top panel) The Zn/Fe for the model KN25-LN18Ka4-S for the best values of $(D_{\rm col}, Z_{\rm eff})$, and contrasting models with $D_{\rm col} = 0$ and 1 respectively. 
    (middle panel) The Zn/Fe for the model KN25-LN18Ka4-S with varying limits to $Z_{\rm eff}$ at which collapsars can form. Limits include 0.025 (black triangle), 0.012 (red star), 0.002 (green square), and 0 (purple cross). (bottom panel) The Zn/Fe for the model KN25-LN18Ka4-S for the best values of $(D_{\rm col}, Z_{\rm eff})$, and contrasting models with $F_{\rm top} = 0$ and 1, respectively. The value of $F_{\rm top}$ varies the percent of 20, 30, and 40 solar mass collapsars by 0.67(1-$F_{\rm top}$), 0.33, 0.67$F_{\rm top}$ for 20, 30, and 40 $M_\odot$ stars.}
    \label{fig:Zn_study}
\end{figure}

In the top panel of Figure \ref{fig:Zn_study}, we plot the contrasting study with one of the GCE model LN25-LN18Ka4-S, but with given $D_{\rm col} = 0$ and 1 for comparison. The $D_{\rm col} = 0$ model corresponds to the best model in Paper II, where no collapsar is considered. Clearly, the trend going everywhere underneath suggests that collapsar is a necessary component in the supernova population, especially in the early universe. The best model is the same model as in Figure \ref{fig:xfe_z_plot}. The $D_{\rm col} = 1$ refers to an extreme case where all massive stars explode as a collapsar. We also include the HMPS sample. The high Zn/Fe of that HMPS suggests that indeed there exists an environment where collapsar is more prominent relative to spherical massive star explosion. However, we also remind that this HMPS could be a result of the local environment instead of a global representation of the stellar population, which does not eliminate the possibility of an ordinary CCSN explosion.

In the middle panel, we compare models with the same $D_{\rm col}$ but different $Z_{\rm off}$, with the same parameter as LN25-LN18Ka4-S. This studies the effect of how the switch-off metallicity, implied from the stellar evolutionary models, affects the Zn production. When $Z_{\rm off} = 0$, this refers to an immediate switch-off of the collapsar channel, which is equivalent to $D_{\rm col} = 0$. For a more standard value for $Z_{\rm off} = 0.002$ ($\sim 0.1 Z_{\odot}$), the Zn/Fe ratio sharply drops beyond that metallicity. The spherical massive star models dilute the Zn/Fe production in the aspherical model. The average value from the stellar data indicates a non-zero contribution beyond $[Fe/H] > -1$. The [Zn/Fe] with $Z_{\rm off} = 0.012, 0.025$ is comparable with the stellar trend. 

In the bottom panel, we compare different $f_{\rm top}$ while fixing $D_{\rm col}$ and $Z_{\rm off}$. This study examines whether the stellar population for the collapsar family is beyond the standard (Salpeter relation),  bottom-heavy, or top-heavy. In our parameter study, top-heavy (bottom-heavy) stands for all collapsars that come from our 40 (20) $M_{\odot}$ model. The Salpeter relation (the best model) is sufficient for reconciling with the galactic trend. For the HMPS, the best value of $D_{\rm col}$ is too low to explain the [Zn/Fe]. A high $f_{\rm top} = 1$ is required to match [Zn/Fe] $\sim 0.8$. On the other hand, the model with a $f_{\rm top} = 0$ is similar to the model $D_{\rm col}$. This is expected because, as seen in previous sections, the lower mass collapsar model does not significantly create much Zn compared to the 40 $M_{\odot}$ models. Thus, a bottom-heavy collapsar population will be less likely to reconcile with the stellar data. Combining with the top plot, this suggests that the HMPS HE 1327-2326 could have originated in a collapsar-dominated environment, or an environment with mixing CCSN and collapsar, but the collapsars are mostly top-heavy in their population.

\begin{figure}
    \centering
    \includegraphics[width=0.98\linewidth]{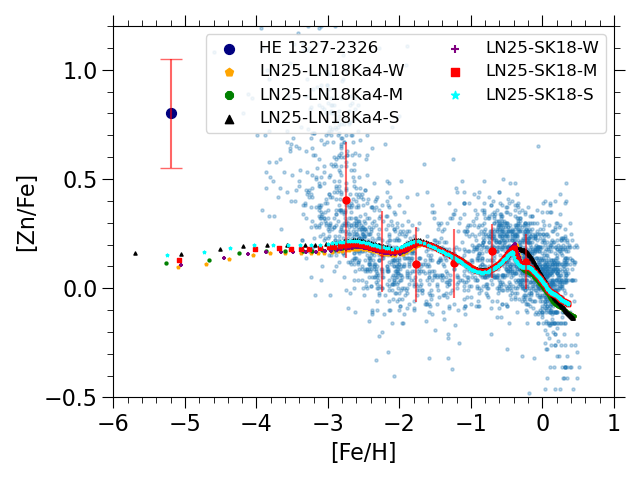}
    \caption{The Zn/Fe for various SN Ia models with our different Collapsar models. The Navy blue circle represents HE 1327-2326, a hyper-metal-poor star extremely dense with zinc. We include LN25-LN18Ka4-S (black triangles), LN25-LN18Ka4-M (green octagon), LN25-LN18Ka4-W (orange pentagon), LN25-SK18-S (cyan star), LN25-SK18-M (red square), and LN25-SK18-W (purple cross). 
    }
    \label{fig:Zn_trend_bestmodels}
\end{figure}

In Figure \ref{fig:Zn_trend_bestmodels} we plot the [Zn/Fe] for the 6 models using the best-fit model parameters. The similarity of the 6 models, regardless of the SN Ia input and the collapsar model, indicates that various explosion energetics models do not result in significant changes to the chemical trends. However, the trends in models following that of the observational data from stars indicate the general importance of aspherical explosions in the universe. 

\subsection{Applications to the Perseus Cluster}

%\begin{table}[]
%    \centering
%    \caption{The best model and the corresponding $D_{\rm col}$ and $Z_{\rm off}$ based on the $\chi^2$ describes the deviation from the abundance bins of the Milky Way Stars. The [Zn/Fe] corresponds 
%    [X/Fe]$= \log_{10}[$($X$/Fe)/($X$/Fe)$_{\odot}$].}
%    \begin{tabular}{c c c c c}
%        \hline
%         Model & $D_{\rm col}$ & $Z_{\rm off}$ & $\chi^2$ & [Zn/Fe]\\ \hline
%         SK18(Strong)(Sum) & 0.2 & 0.01175 & 23.279 \\
%         Ka4(Strong)(Prod) & 0.2 & 0.012 & 54.376 \\
%         & \\
%         & \\
%         & \\
%         & \\ \hline
%    \end{tabular}
%    \label{tab:best_model_SAGA}
%\end{table}

\begin{figure}
    \centering
    \includegraphics[width=0.98\linewidth]{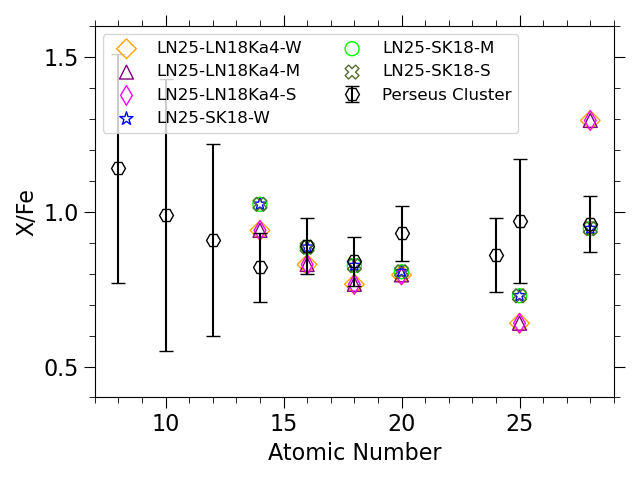}
    \caption{The chemical abundances X/Fe for various collapsar models. The included models are LN25-LN18Ka4-S (pink diamond), LN25-LN18Ka4-M (purple triangle), LN25-LN18Ka4-W (orange diamond), LN25-SK18-S (green X), LN25-SK18-M (lime circle), LN25-SK18-W (blue star). The error bars correspond to the measurements of the Perseus Cluster from \cite{Simionescu2019Perseus}.}
    \label{fig:perseus_comarpison}
\end{figure}

In Figure \ref{fig:perseus_comarpison}, we compare the chemical abundance of the best-fit models at the current universe from the GCE models with those measured in the Perseus Cluster. The best models are based on the LN25 CCSN model, the 2 best SN Ia models from Paper II (LN18(Ka4) and SK18), and the 3 collapsar energetics models. We then surveyed the parameters $D_{col}$ and $f_{top}$ to find the best-fit. The parameters for these best-fit models are shown in Table \ref{tab:best_model_Perseus}. The level of fitting does not depend strongly on the collapsar jet energetics.  
On the other hand, the proximity of the current chemical abundance from the GCE model is strongly affected by the input SN Ia models. This agrees with Figure \ref{fig:Zn_trend_bestmodels} on the Zn/Fe trend. 
%Our best-fit models for the Strong and Medium jet models indicate a bottom-heavy collapsar population. The Weak energetic model produced a mass-independent top-heavy collapsar population. However, similar to \ref{fig:Zn_trend_bestmodels}, we see no significant difference between the different collapsar energetic models based on best-fit parameters. 

As noted in Figure \ref{fig:xfe_z_plot}, all best models have a similar trend once the ordinary massive stars and SNe Ia dominate the synthesis of Fe-group elements.  

\section{Discussion}
\label{sec:discussion}

\subsection{Relative Importance of Collapsar}

The collapsar model has been independently investigated by the lens of chemical elements and the long GRB population. These studies provide further indication of the typical metallicity range for a collapsar in the cosmological context. 

In \cite{Bignone2017GRBzLim}, the long GRB host population collected from catalogues including \texttt{BAT6} \citep{Salvaterra2012GRBCat}, \texttt{TOUGH} \citep{Hjorth2012GRBCat} and \texttt{SHOALS} \citep{Perley2016GRBCat} is compared with the cosmological N-body simulation \texttt{Illustris}. The mass distribution of the host galaxy of the long GRB events suggests a metallicity cutoff between 0.3 -- 0.6. 

Besides Zn, the r-process elements [Eu/Fe] are another widely studied element ratio for constraining. Eu is conventionally attributed to the neutron star mergers \citep{Burbidge1957SNReview, Cameron1957HeavyEle}. However, there exists a delay time in the merger process by the formation gravitational wave timescale for Eu to be formed by this channel. Thus, early [Eu/Fe] found before [Fe/H]<-2.5 has likely other sources. In \cite{Brauer2021EuFeCollapsar} collapsar is shown to be a robust source of Eu in the early universe. Later study shows that collapsars need to have a switch-off metallicity based on the composition of the cold gas in the Galaxy \citep{Fraser2022CollapsarGCE}. A similar threshold of 0.3 $Z_{\odot}$ is obtained. 

These constraints further agree with the $Z_{\rm off} \sim 0.5$ based on our multi-elemental constraints, which further confirm its indispensable role in chemical enrichment in the early universe. 

\subsection{Conclusion}

In this article, we have presented the new jet-driven supernova (collapsar) models for 20, 30, and 40 $M_{\odot}$ with different jet energetics. We show that the new aspherical stellar explosion models result in a distinctive chemical abundance pattern from the previous works, which highlights the importance of the massive star input. The collapsar model yields features pronounced by lower-mass elements (e.g., Mn, Si) and very significant Ni and Zn. For high mass models, the production of lower Fe-group elements is also pronounced. This provides additional variation to the canonical spherical massive star explosion models, which could not fully match some recently observed metal-poor galaxies \citep[see e.g., ][]{Leung2024Jet2}.

We show that the new collapsar models could also match the observed Si, S, Ar, and Ca ratios in the Perseus Cluster. We compared the chemical yields with some well-measured metal-poor stars. However, the observed high Ca and low Cr abundances in these stars indicate some additional supernova environment. The collapsar models can contribute to the $^{56}$Ni and ejecta mass for supernovae reported in transient surveys (e.g., ZTF) except for those with high ejecta mass but low $^{56}$Ni. The new models also show trends of  [V/Fe] and [Ti/Fe] with a slope and magnitude agreeing with the metal-poor stars from surveys.   

By using the new collapsar models in the Galactic Chemical Evolution as the additional stellar input, we show that regardless of the different SN Ia yields and the actual collapsar energetic, a nonzero $D_{\rm col} \approx 0.2 - 0.3$ is necessary to match the elemental trends based on Milky Way stars. A too high value of $D_{\rm col}$ results in a too low production to match the galactic stars. It further indicates a sub-solar metallicity limit ($Z_{\rm off} \approx 0.012$) beyond which the collapsar becomes less important. A similar result is that a too low $Z_{\rm off}$ will also lead to a mismatch with the galactic stars. The limits in the metallicity could correlate with the metallicity threshold observed for some long GRBs \citep{Bignone2017GRBzLim}.  

We pay attention to the Zn production in our control study. A non-zero $D_{\rm col}$ is necessary not only for typical Si-group and Fe-group elements, but also for the magnitude and trend for [Zn/Fe] measured in Milky Way stars. The highly Zn-enriched HMPS HE 1327-2326 favours the collapsar channel, attributed by either (1) a collapsar-dominated environment, or (2) a mixed collapsar + CCSN environment, but dominated by 40 $M_{\odot}$ (top-heavy) collapsars. Further discoveries of low-metallicity, high Zn/Fe stars could provide us with further constraints on the relative rate of collapsars over CCSNe in the early universe.  

We also search for the best model that could explain the Perseus Cluster. However, the high metallicity environment results in other supernova classes that shadow the contribution of the collapsar. Thus, the Perseus Cluster itself does not lead to explicit constraints on the collapsar parameters. 

This article demonstrates the possibility of including the collapsar channel for a better match of chemical elements. This shows that collapsar is not only an extension of massive star explosion from spherical to aspherical mode, but also a distinctive component required to match the various chemical trends in stellar and galactic systems presented in this article. The future derived rate of long GRB, metallicity dependence of stars forming a collapsar, and more Zn-enriched early stellar objects will provide a more reliable indication of the fraction of collapsars in cosmic history. This provides indirect constraints on the stellar evolutionary model of the mass loss and angular momentum transfer in low-metallicity stars.

\software{  Numpy \citep{Numpy},
            Matplotlib \citep{Matplotlib},
            Pandas \citep{Pandas}
          }

\section*{Acknowledgment}

We thank Frank Timmes for the open-source subroutines of the Helmholtz equation of state and the Torch nuclear reaction network. 
This material is based upon work supported by the National Science Foundation under Grant AST-2316807.
K.N. acknowledges support by the World Premier International Research Center Initiative (WPI), and JSPS KAKENHI Grant Numbers JP20K04024, JP21H04499, JP23K03452, and JP25K01046. 
A.S. acknowledges the Kavli IPMU for the continued hospitality. SRON Netherlands Institute for Space Research is supported financially by NWO.

\vspace{5mm}

\newpage
\appendix

\section{Yield Table of the Characteristic Jet-Driven Supernovae}

In the main text, we have presented the dynamics and general nucleosynthetic features of the jet-driven supernova models using the new massive star models as progenitors. Here we provide the table form of the supernova ejecta masses for models using the default jet parameters in Table \ref{table:yield_isotopes} for the stable isotopes, Table \ref{table:unstable_isotopes} for the unstable short-lived isotopes, and Table \ref{table:yield_elements} for the elements \ref{table:yield_elements}. 
%For the comprehensive tables, we refer interested readers to the online version on Zenodo (\href{Link}{10.5281/zenodo.18449526}). 

\begin{longtable}[h]{c c c c}
 \caption{The isotope yields table for spherical explosion models. Masses are in units of $M_{\odot}$ and energies are in unit of $10^{51}$ erg.\label{table:yield_isotopes}} \\
 %\centering
 
%set the format of the table
 %\hline 
 %\multicolumn{8}{c}{Table \ref{table:yield_isotopes}: Isotope yield table for supernova models.}  \\
 %\hline
 Isotope & M20-100-100 & M30-100-100 & M40-100-100 \\
 \hline
 \endfirsthead

 %\hline
 \multicolumn{4}{c}{\textit{Continuation of Table \ref{table:yield_isotopes}.}} \\
 %\hline
 Isotope & M20-100-100 & M30-100-100 & M40-100-100 \\
 \hline
 \endhead

 \hline
 \endfoot

 %\hline
 %\multicolumn{8}{ c }{\textit{End of Table \ref{table:yield_isotopes}.}} \\
 %\hline
 %Isotope & M15A22a003 & M20A22a003 & M25A22a003 & M40A22a003 & Chand & sub-Chand \\
 \hline
 \endlastfoot

 $^{12}$C & $2.55 \times 10^{-1}$ & $7.51 \times 10^{-1}$ & $5.81 \times 10^{-1}$ \\
 $^{13}$C & $1.43 \times 10^{-3}$ & $1.73 \times 10^{-6}$ & $4.49 \times 10^{-4}$ \\
 $^{14}$N & $2.17 \times 10^{-2}$ & $1.3 \times 10^{-2}$ & $1.9 \times 10^{-3}$ \\
 $^{15}$N & $6.59 \times 10^{-6}$ & $3.21 \times 10^{-6}$ & $4.4 \times 10^{-7}$ \\
 $^{16}$O & $18.85 \times 10^{-1}$ & $34.28 \times 10^{-1}$ & $42.35 \times 10^{-1}$ \\
 $^{17}$O & $7.84 \times 10^{-3}$ & $1.43 \times 10^{-3}$ & $9.17 \times 10^{-4}$ \\
 $^{18}$O & $2.80 \times 10^{-7}$ & $3.75 \times 10^{-5}$ & $6.22 \times 10^{-7}$ \\
 $^{19}$F & $1.33 \times 10^{-6}$ & $1.8 \times 10^{-5}$ & $2.27 \times 10^{-7}$ \\
 $^{20}$Ne & $6.78 \times 10^{-1}$ & $7.15 \times 10^{-1}$ & $7.62 \times 10^{-1}$ \\
 $^{21}$Ne & $5.24 \times 10^{-4}$ & $4.46 \times 10^{-3}$ & $2.42 \times 10^{-3}$ \\
 $^{22}$Ne & $4.20 \times 10^{-3}$ & $1.46 \times 10^{-2}$ & $4.34 \times 10^{-3}$ \\
 $^{23}$Na & $4.49 \times 10^{-3}$ & $1.57 \times 10^{-2}$ & $1.43 \times 10^{-2}$ \\
 $^{24}$Mg & $1.95 \times 10^{-1}$ & $8.4 \times 10^{-2}$ & $8.43 \times 10^{-2}$ \\
 $^{25}$Mg & $2.1 \times 10^{-2}$ & $2.7 \times 10^{-2}$ & $2.28 \times 10^{-2}$ \\
 $^{26}$Mg & $1.87 \times 10^{-2}$ & $2.71 \times 10^{-2}$ & $2.79 \times 10^{-2}$ \\
 $^{26}$Al & $1.41 \times 10^{-28}$ & $1.90 \times 10^{-28}$ & $1.65 \times 10^{-28}$ \\
 $^{27}$Al & $2.2 \times 10^{-2}$ & $1.41 \times 10^{-2}$ & $1.39 \times 10^{-2}$ \\
 $^{28}$Si & $1.28 \times 10^{-1}$ & $8.67 \times 10^{-2}$ & $3.85 \times 10^{-2}$ \\
 $^{29}$Si & $9.61 \times 10^{-3}$ & $2.13 \times 10^{-3}$ & $1.76 \times 10^{-3}$ \\
 $^{30}$Si & $1.11 \times 10^{-2}$ & $5.47 \times 10^{-3}$ & $2.77 \times 10^{-3}$ \\
 $^{31}$P & $1.46 \times 10^{-3}$ & $7.25 \times 10^{-4}$ & $4.39 \times 10^{-4}$ \\
 $^{32}$S & $3.93 \times 10^{-2}$ & $4.5 \times 10^{-2}$ & $1.95 \times 10^{-2}$ \\
 $^{33}$S & $3.54 \times 10^{-4}$ & $2.17 \times 10^{-4}$ & $9.51 \times 10^{-5}$ \\
 $^{34}$S & $2.17 \times 10^{-3}$ & $1.80 \times 10^{-3}$ & $6.33 \times 10^{-4}$ \\
 $^{36}$S & $4.26 \times 10^{-7}$ & $2.59 \times 10^{-7}$ & $9.60 \times 10^{-8}$ \\
 $^{35}$Cl & $1.41 \times 10^{-4}$ & $1.27 \times 10^{-4}$ & $6.82 \times 10^{-5}$ \\
 $^{37}$Cl & $1.48 \times 10^{-5}$ & $1.40 \times 10^{-5}$ & $1.16 \times 10^{-5}$ \\
 $^{36}$Ar & $6.22 \times 10^{-3}$ & $7.25 \times 10^{-3}$ & $4.3 \times 10^{-3}$ \\
 $^{38}$Ar & $6.57 \times 10^{-4}$ & $7.52 \times 10^{-4}$ & $3.42 \times 10^{-4}$ \\
 $^{40}$Ar & $5.69 \times 10^{-9}$ & $4.59 \times 10^{-9}$ & $1.37 \times 10^{-8}$ \\
 $^{39}$K & $4.64 \times 10^{-5}$ & $3.90 \times 10^{-5}$ & $2.74 \times 10^{-5}$ \\
 $^{40}$K & $4.91 \times 10^{-8}$ & $5.58 \times 10^{-8}$ & $5.54 \times 10^{-8}$ \\
 $^{41}$K & $3.29 \times 10^{-6}$ & $3.6 \times 10^{-6}$ & $2.16 \times 10^{-6}$ \\
 $^{40}$Ca & $5.17 \times 10^{-3}$ & $6.39 \times 10^{-3}$ & $3.81 \times 10^{-3}$ \\
 $^{42}$Ca & $1.80 \times 10^{-5}$ & $1.96 \times 10^{-5}$ & $2.79 \times 10^{-5}$ \\
 $^{43}$Ca & $6.75 \times 10^{-7}$ & $2.19 \times 10^{-6}$ & $3.29 \times 10^{-6}$ \\
 $^{44}$Ca & $1.74 \times 10^{-5}$ & $5.85 \times 10^{-5}$ & $2.7 \times 10^{-5}$ \\
 $^{46}$Ca & $8.30 \times 10^{-11}$ & $2.20 \times 10^{-10}$ & $5.26 \times 10^{-7}$ \\
 $^{48}$Ca & $6.47 \times 10^{-15}$ & $9.65 \times 10^{-14}$ & $5.98 \times 10^{-9}$ \\
 $^{45}$Sc & $2.17 \times 10^{-7}$ & $3.18 \times 10^{-7}$ & $3.81 \times 10^{-6}$ \\
 $^{46}$Ti & $9.21 \times 10^{-6}$ & $7.21 \times 10^{-6}$ & $2.80 \times 10^{-5}$ \\
 $^{47}$Ti & $3.92 \times 10^{-6}$ & $1.44 \times 10^{-5}$ & $2.4 \times 10^{-5}$ \\
 $^{48}$Ti & $1.47 \times 10^{-4}$ & $4.15 \times 10^{-4}$ & $1.63 \times 10^{-4}$ \\
 $^{49}$Ti & $8.31 \times 10^{-6}$ & $9.96 \times 10^{-6}$ & $3.88 \times 10^{-5}$ \\
 $^{50}$Ti & $4.99 \times 10^{-10}$ & $2.74 \times 10^{-10}$ & $2.28 \times 10^{-5}$ \\
 $^{50}$V & $1.28 \times 10^{-9}$ & $1.16 \times 10^{-9}$ & $2.52 \times 10^{-6}$ \\
 $^{51}$V & $2.12 \times 10^{-5}$ & $2.75 \times 10^{-5}$ & $1.4 \times 10^{-4}$ \\
 $^{50}$Cr & $7.12 \times 10^{-5}$ & $4.54 \times 10^{-5}$ & $1.91 \times 10^{-4}$ \\
 $^{52}$Cr & $9.79 \times 10^{-4}$ & $1.86 \times 10^{-3}$ & $1.15 \times 10^{-3}$ \\
 $^{53}$Cr & $1.25 \times 10^{-4}$ & $1.47 \times 10^{-4}$ & $3.68 \times 10^{-4}$ \\
 $^{54}$Cr & $4.1 \times 10^{-8}$ & $1.75 \times 10^{-8}$ & $9.16 \times 10^{-5}$ \\
 $^{55}$Mn & $7.76 \times 10^{-4}$ & $6.50 \times 10^{-4}$ & $1.6 \times 10^{-3}$ \\
 $^{54}$Fe & $5.0 \times 10^{-3}$ & $3.42 \times 10^{-3}$ & $8.99 \times 10^{-3}$ \\
 $^{56}$Fe & $5.94 \times 10^{-2}$ & $1.10 \times 10^{-1}$ & $5.30 \times 10^{-2}$ \\
 $^{57}$Fe & $3.53 \times 10^{-3}$ & $6.31 \times 10^{-3}$ & $4.3 \times 10^{-3}$ \\
 $^{58}$Fe & $1.40 \times 10^{-6}$ & $2.56 \times 10^{-8}$ & $2.51 \times 10^{-4}$ \\
 $^{60}$Fe & $1.31 \times 10^{-11}$ & $1.32 \times 10^{-18}$ & $2.1 \times 10^{-6}$ \\
 $^{59}$Co & $2.60 \times 10^{-4}$ & $3.66 \times 10^{-4}$ & $1.25 \times 10^{-3}$ \\
 $^{58}$Ni & $1.54 \times 10^{-2}$ & $1.35 \times 10^{-2}$ & $2.98 \times 10^{-2}$ \\
 $^{60}$Ni & $1.97 \times 10^{-3}$ & $2.18 \times 10^{-3}$ & $1.33 \times 10^{-2}$ \\
 $^{61}$Ni & $1.29 \times 10^{-4}$ & $2.6 \times 10^{-4}$ & $7.1 \times 10^{-4}$ \\
 $^{62}$Ni & $1.10 \times 10^{-3}$ & $1.56 \times 10^{-3}$ & $3.78 \times 10^{-3}$ \\
 $^{64}$Ni & $2.29 \times 10^{-6}$ & $3.15 \times 10^{-10}$ & $5.22 \times 10^{-5}$ \\
 $^{63}$Cu & $3.78 \times 10^{-5}$ & $1.60 \times 10^{-5}$ & $6.28 \times 10^{-4}$ \\
 $^{65}$Cu & $9.19 \times 10^{-6}$ & $4.99 \times 10^{-6}$ & $2.27 \times 10^{-4}$ \\
 $^{64}$Zn & $7.83 \times 10^{-5}$ & $1.37 \times 10^{-4}$ & $9.93 \times 10^{-3}$ \\
 $^{66}$Zn & $1.42 \times 10^{-4}$ & $2.72 \times 10^{-5}$ & $8.96 \times 10^{-4}$ \\
 $^{67}$Zn & $2.32 \times 10^{-6}$ & $5.74 \times 10^{-7}$ & $9.74 \times 10^{-5}$ \\
 $^{68}$Zn & $4.10 \times 10^{-6}$ & $8.31 \times 10^{-6}$ & $9.65 \times 10^{-4}$ \\
 $^{70}$Zn & $3.32 \times 10^{-9}$ & $9.89 \times 10^{-17}$ & $1.99 \times 10^{-7}$ \\

\end{longtable}

\begin{longtable}[c]{c c c c c c c}
 \caption{Same as Table \ref{table:yield_isotopes} but for short-lived radioactive isotopes. \label{table:unstable_isotopes}} \\
 %\centering
 
%set the format of the table
 %\hline 
 %\multicolumn{8}{c}{Table \ref{table:yield_isotopes}: Isotope yield table for supernova models.}  \\
 %\hline
 Isotope & M20-100-100 & M30-100-100 & M40-100-100 \\
 \hline
 \endfirsthead

 %\hline
 \multicolumn{4}{c}{\textit{Continuation of Table \ref{table:unstable_isotopes}.}} \\
 %\hline
 Isotope & M20-100-100 & M30-100-100 & M40-100-100 \\
 \hline
 \endhead

 \hline
 \endfoot

 %\hline
 %\multicolumn{8}{ c }{\textit{End of Table \ref{table:yield_isotopes}.}} \\
 %\hline
 %Isotope & M15A22a003 & M20A22a003 & M25A22a003 & M40A22a003 & Chand & sub-Chand \\
 \hline
 \endlastfoot

  $^{22}$Na & $8.94 \times 10^{-6}$ & $3.92 \times 10^{-6}$ & $8.25 \times 10^{-6}$ \\
 $^{26}$Al & $4.62 \times 10^{-5}$ & $2.54 \times 10^{-5}$ & $9.68 \times 10^{-5}$ \\
 $^{39}$Ar & $2.81 \times 10^{-8}$ & $1.47 \times 10^{-8}$ & $4.91 \times 10^{-8}$ \\
 $^{40}$K & $4.94 \times 10^{-8}$ & $5.61 \times 10^{-8}$ & $5.57 \times 10^{-8}$ \\
 $^{41}$Ca & $3.14 \times 10^{-6}$ & $2.86 \times 10^{-6}$ & $2.14 \times 10^{-6}$ \\
 $^{44}$Ti & $1.61 \times 10^{-5}$ & $5.45 \times 10^{-5}$ & $1.19 \times 10^{-5}$ \\
 $^{48}$V & $2.26 \times 10^{-8}$ & $2.42 \times 10^{-8}$ & $2.96 \times 10^{-6}$ \\
 $^{49}$V & $5.40 \times 10^{-8}$ & $6.52 \times 10^{-8}$ & $1.69 \times 10^{-5}$ \\
 $^{53}$Mn & $7.29 \times 10^{-6}$ & $3.6 \times 10^{-6}$ & $2.35 \times 10^{-4}$ \\
 $^{60}$Fe & $1.94 \times 10^{-10}$ & $2.8 \times 10^{-17}$ & $2.88 \times 10^{-5}$ \\
 $^{56}$Co & $9.68 \times 10^{-6}$ & $3.22 \times 10^{-6}$ & $5.32 \times 10^{-5}$ \\
 $^{57}$Co & $1.99 \times 10^{-5}$ & $3.68 \times 10^{-6}$ & $2.37 \times 10^{-4}$ \\
 $^{60}$Co & $4.37 \times 10^{-8}$ & $2.56 \times 10^{-12}$ & $1.93 \times 10^{-5}$ \\
 $^{56}$Ni & $5.93 \times 10^{-2}$ & $1.10 \times 10^{-1}$ & $5.21 \times 10^{-2}$ \\
 $^{57}$Ni & $3.51 \times 10^{-3}$ & $6.31 \times 10^{-3}$ & $3.72 \times 10^{-3}$ \\
 $^{59}$Ni & $7.76 \times 10^{-5}$ & $5.90 \times 10^{-5}$ & $9.93 \times 10^{-4}$ \\
 $^{63}$Ni & $1.94 \times 10^{-6}$ & $8.84 \times 10^{-12}$ & $2.11 \times 10^{-5}$ \\

 \end{longtable}

 \begin{longtable}[c]{c c c c c c c}
 \caption{Same as \ref{table:yield_isotopes} but for the elemental yields. \label{table:yield_elements}} \\
 %\centering
 
%set the format of the table
 %\hline 
 %\multicolumn{8}{c}{Table \ref{table:yield_isotopes}: Isotope yield table for supernova models.}  \\
 %\hline
 Isotope & M20-100-100 & M30-100-100 & M40-100-100 \\
 \hline
 \endfirsthead

 %\hline
 \multicolumn{4}{c}{\textit{Continuation of Table \ref{table:yield_elements}.}} \\
 %\hline
 Isotope & M20-100-100 & M30-100-100 & M40-100-100 \\
 \hline
 \endhead

 \hline
 \endfoot

 %\hline
 %\multicolumn{8}{ c }{\textit{End of Table \ref{table:yield_isotopes}.}} \\
 %\hline
 %Isotope & M15A22a003 & M20A22a003 & M25A22a003 & M40A22a003 & Chand & sub-Chand \\
 \hline
 \endlastfoot

  C & $2.56 \times 10^{-1}$ & $7.51 \times 10^{-1}$ & $5.82 \times 10^{-1}$ \\
 N & $2.17 \times 10^{-2}$ & $1.3 \times 10^{-2}$ & $1.9 \times 10^{-3}$ \\
 O & $18.92 \times 10^{-1}$ & $34.29 \times 10^{-1}$ & $42.36 \times 10^{-1}$ \\
 F & $1.33 \times 10^{-6}$ & $1.8 \times 10^{-5}$ & $2.27 \times 10^{-7}$ \\
 Ne & $6.82 \times 10^{-1}$ & $7.34 \times 10^{-1}$ & $7.69 \times 10^{-1}$ \\
 Na & $4.49 \times 10^{-3}$ & $1.57 \times 10^{-2}$ & $1.43 \times 10^{-2}$ \\
 Mg & $2.34 \times 10^{-1}$ & $1.28 \times 10^{-1}$ & $1.35 \times 10^{-1}$ \\
 Al & $2.2 \times 10^{-2}$ & $1.41 \times 10^{-2}$ & $1.39 \times 10^{-2}$ \\
 Si & $1.49 \times 10^{-1}$ & $9.43 \times 10^{-2}$ & $4.31 \times 10^{-2}$ \\
 P & $1.46 \times 10^{-3}$ & $7.25 \times 10^{-4}$ & $4.39 \times 10^{-4}$ \\
 S & $4.18 \times 10^{-2}$ & $4.26 \times 10^{-2}$ & $2.3 \times 10^{-2}$ \\
 Cl & $1.56 \times 10^{-4}$ & $1.41 \times 10^{-4}$ & $7.98 \times 10^{-5}$ \\
 Ar & $6.88 \times 10^{-3}$ & $8.0 \times 10^{-3}$ & $4.37 \times 10^{-3}$ \\
 K & $4.97 \times 10^{-5}$ & $4.21 \times 10^{-5}$ & $2.97 \times 10^{-5}$ \\
 Ca & $5.20 \times 10^{-3}$ & $6.47 \times 10^{-3}$ & $3.87 \times 10^{-3}$ \\
 Sc & $2.17 \times 10^{-7}$ & $3.18 \times 10^{-7}$ & $3.81 \times 10^{-6}$ \\
 Ti & $1.68 \times 10^{-4}$ & $4.47 \times 10^{-4}$ & $2.74 \times 10^{-4}$ \\
 V & $2.12 \times 10^{-5}$ & $2.75 \times 10^{-5}$ & $1.7 \times 10^{-4}$ \\
 Cr & $1.17 \times 10^{-3}$ & $2.5 \times 10^{-3}$ & $1.81 \times 10^{-3}$ \\
 Mn & $7.76 \times 10^{-4}$ & $6.50 \times 10^{-4}$ & $1.6 \times 10^{-3}$ \\
 Fe & $6.79 \times 10^{-2}$ & $1.19 \times 10^{-1}$ & $6.62 \times 10^{-2}$ \\
 Co & $2.60 \times 10^{-4}$ & $3.66 \times 10^{-4}$ & $1.25 \times 10^{-3}$ \\
 Ni & $1.86 \times 10^{-2}$ & $1.75 \times 10^{-2}$ & $4.76 \times 10^{-2}$ \\
 Cu & $4.70 \times 10^{-5}$ & $2.10 \times 10^{-5}$ & $8.55 \times 10^{-4}$ \\
 Zn & $2.27 \times 10^{-4}$ & $1.73 \times 10^{-4}$ & $1.18 \times 10^{-2}$ \\

\end{longtable}

\bibliographystyle{aasjournal}
\pagestyle{plain}
\bibliography{biblio}

\end{document}